\documentclass[review, 11pt,sort&compress]{elsarticle}
\usepackage[utf8]{inputenc}
\usepackage[toc,page]{appendix}
\usepackage{amssymb}            
\usepackage{babel}
\usepackage{graphicx}
\usepackage{booktabs}
\usepackage{xcolor}
\usepackage{gensymb} 
\usepackage{amsmath,amsfonts,amsthm,bm,mathtools} 
\usepackage{mathrsfs} 
\usepackage{geometry}
\usepackage{gensymb}
\usepackage{multirow}
\usepackage[skip=6pt plus1pt, indent=40pt]{parskip}
\usepackage{setspace} 

\usepackage{multicol}
\usepackage{mdframed}
\usepackage{nomencl}
\makenomenclature
\usepackage{etoolbox}
\renewcommand\nomgroup[1]{%
  \item[\bfseries
  \ifstrequal{#1}{A}{Greek letters}{%
  \ifstrequal{#1}{B}{Lowercase letters}{%
   \ifstrequal{#1}{C}{Uppercase letters}{%
  \ifstrequal{#1}{D}{Others}{}}}}%
]}

\journal{Modelling and Simulation in Materials Science and Engineering}
\usepackage{graphicx}
\graphicspath{{Figures/}}
\usepackage{caption}
\usepackage{subcaption}
\usepackage{float}
\usepackage{hyperref}
\hypersetup{colorlinks=true, linkcolor=blue, citecolor=black, filecolor=magenta, urlcolor=black}
\usepackage{soul}
\usepackage{accents}
\usepackage{color,soul} 
\usepackage{color} 
\usepackage{xcolor,soul}
\sethlcolor{lightgray}

\begin{document}
\begin{frontmatter}
\title{A nonlinear phase-field model of corrosion with charging kinetics of electric double layer}

\author[1]{Maciej Makuch}
\author[2]{Sasa Kovacevic}
\author[1]{Mark R. Wenman}
\author[2]{Emilio Mart\'inez-Pa\~neda \corref{cor1}}
\ead{emilio.martinez-paneda@eng.ox.ac.uk}

\cortext[cor1]{Corresponding authors}

\address[1]{Imperial College London, Centre for Nuclear Engineering, South Kensington Campus, London SW7 2AZ, UK}
\address[2]{Department of Engineering Science, University of Oxford, Oxford OX1 3PJ, UK}

\begin{abstract}
A nonlinear phase-field model is developed to simulate corrosion damage. The motion of the electrode$-$ electrolyte interface follows the usual kinetic rate theory for chemical reactions based on the Butler$-$ Volmer equation. The model links the surface polarization variation associated with the charging kinetics of an electric double layer (EDL) to the mesoscale transport. The effects of the EDL are integrated as a boundary condition on the solution potential equation. The boundary condition controls the magnitude of the solution potential at the electrode$-$electrolyte interface. The ion concentration field outside the EDL is obtained by solving the electro$-$diffusion equation and Ohm's law for the solution potential. The model is validated against the classic benchmark pencil electrode test. The framework developed reproduces experimental measurements of both pit kinetics and transient current density response. The model enables more accurate information on corrosion damage, current density, and environmental response in terms of the distribution of electric potential and charged species. The sensitivity analysis for different properties of the EDL is performed to investigate their role in the electrochemical response of the system. Simulation results show that the properties of the EDL significantly influence the transport of ionic species in the electrolyte.\\
\end{abstract}
\end{frontmatter}
\cleardoublepage
\nomenclature[A,1]{$\alpha$}{Anodic charge transfer coefficient}
\nomenclature[A,2]{$\zeta$}{Electric potential source term}
\nomenclature[A,3]{$\eta$}{Overpotential}
\nomenclature[A,4]{$\kappa$}{Gradient energy coefficient}
\nomenclature[A,5]{$\lambda$}{Effective electric conductivity}
\nomenclature[A,6]{$\lambda^l$}{Electric conductivity of the liquid domain}
\nomenclature[A,7]{$\lambda^l$}{Electric conductivity of the solid domain}
\nomenclature[A,8]{$\mu_i^{\theta}$}{Reference chemical potential}
\nomenclature[A,9]{$\xi$}{Geometrical factor}
\nomenclature[A,10]{$\phi$}{Phase-field variable}
\nomenclature[A,11]{$\psi_l$}{Electric potential}
\nomenclature[A,12]{$\psi_l^{dl}$}{Electric potential outside EDL}
\nomenclature[A,13]{$\psi_l^{ref}$}{Reference electric potential}
\nomenclature[A,14]{$\psi^0$}{Initial surface polarization}
\nomenclature[A,15]{$\chi$}{Ratio between EDL resistance and electrolyte resistance}
\nomenclature[A,16]{$\omega$}{Height of the double well potential}
\nomenclature[A,17]{$\Gamma$}{Interfacial energy}
\nomenclature[A,18]{$\Omega$}{Domain investigated}
\nomenclature[B,1]{$\overrightarrow{c}$}{Set of ionic concentrations}
\nomenclature[B,2]{$c_{i}$}{Concentration of ionic species $i$}
\nomenclature[B,3]{$\bar{c}_i$}{Normalized ionic concentration of component $i$}
\nomenclature[B,4]{$\bar{c}_{1}^{l}$}{Normalized concentration of metal ions in the liquid phase}
\nomenclature[B,5]{$\bar{c}_{1}^{s}$}{Normalized concentration of metal ions in the solid phase}
\nomenclature[B,6]{$c_{1}^{l,eq}$}{Equilibrium concentration of metal ions in the liquid phase}
\nomenclature[B,7]{$c_{1}^{s,eq}$}{Equilibrium concentration of metal ions in the solid phase}
\nomenclature[B,8]{$\bar{c}_{1}^{l,eq}$}{Normalized equilibrium concentration of metal ions in the liquid phase}
\nomenclature[B,9]{$\bar{c}_{1}^{s,eq}$}{Normalized equilibrium concentration of metal ions in the solid phase}
\nomenclature[B,10]{$f^{chem}$}{Chemical free energy density}
\nomenclature[B,11]{$f^{elec}$}{Electric free energy density}
\nomenclature[B,12]{$f^{grad}$}{Interfacial energy density}
\nomenclature[B,13]{$f_{i}^{chem}$}{Chemical free energy density of component $i$}
\nomenclature[B,14]{$f_{l}^{chem}$}{Chemical free energy density of the liquid phase}
\nomenclature[B,15]{$f_{s}^{chem}$}{Chemical free energy density of the solid phase}
\nomenclature[B,16]{$g(\phi)$}{Double well potential shape function}
\nomenclature[B,17]{$h(\phi)$}{Interpolation function}
\nomenclature[B,18]{$h'(\phi)$}{Derivative of interpolation function with respect to phase-field parameter}
\nomenclature[B,19]{$i_a$}{Anodic current density}
\nomenclature[B,20]{$i_0$}{Exchange current density}
\nomenclature[B,21]{$k_{1b}$}{Backward reaction constant for primary hydrolysis}
\nomenclature[B,22]{$k_{1f}$}{Forward reaction constant for primary hydrolysis}
\nomenclature[B,23]{$k_{2b}$}{Backward reaction constant for water dissociation}
\nomenclature[B,24]{$k_{2f}$}{Forward reaction constant for water dissociation}
\nomenclature[B,25]{$t_c$}{Half-time capacitor charging constant}
\nomenclature[B,26]{$z_i$}{Charge number of component $i$}
\nomenclature[C,1]{$A$}{Free energy density curvature parameter}
\nomenclature[C,2]{$C_{dl}$}{Electric double layer capacitance}
\nomenclature[C,3]{$D_i$}{Effective diffusion coefficient of component $i$}
\nomenclature[C,4]{$D_{i}^{l}$}{Diffusion coefficient of component $i$ in liquid phase}
\nomenclature[C,5]{$D_{i}^{s}$}{Diffusion coefficient of component $i$ in solid phase}
\nomenclature[C,6]{$E_{app}$}{Applied potential difference with respect to reference electrode}
\nomenclature[C,7]{$E_{eq}$}{Equilibrium corrosion potential with respect to reference electrode}
\nomenclature[C,8]{$F$}{Faraday's constant}
\nomenclature[C,9]{$\mathbf{J}_i$}{Electrochemical flux of component $i$}
\nomenclature[C,10]{$K_1$}{Equilibrium constant for primary hydrolysis reaction}
\nomenclature[C,11]{$K_2$}{Equilibrium constant for water dissociation}
\nomenclature[C,12]{$L$}{Enhanced phase-field mobility parameter}
\nomenclature[C,13]{$L_0$}{Constant interfacial mobility parameter}
\nomenclature[C,14]{$R$}{Gas constant}
\nomenclature[C,15]{$R_{dl}$}{Electric double layer resistance}
\nomenclature[C,16]{$R_i$}{Volumetric reaction rate of component $i$}
\nomenclature[C,17]{$R_l$}{Electrolyte resistance}
\nomenclature[C,18]{$T$}{Absolute temperature}
\nomenclature[C,19]{$V_m$}{Molar volume of metal}
\nomenclature[D,1]{$\ell$}{Interface thickness}
\nomenclature[D,2]{$\partial \Omega$}{Domain boundary}
\nomenclature[D,3]{$\partial \Omega_l$}{Domain boundary of electrolyte}
\nomenclature[D,4]{$\partial \Omega_m$}{Domain boundary of metal}
\nomenclature[D,5]{$\mathscr{F}$}{Free energy functional}
\nomenclature[D,6]{EDL}{Electric double layer}
\nomenclature[D,7]{ECM}{Equivalent circuit model}
\nomenclature[D,7]{SCE}{Saturated calomel electrode}
\begin{mdframed}
\small
\begin{singlespace}
\begin{multicols}{2}
\printnomenclature
\end{multicols}
\end{singlespace}
\end{mdframed}
\section{Introduction} \label{sec1}

Pitting corrosion is a particular form of localized corrosion that occurs after a breakdown of the passive film on a metal surface \cite{Liu2023}. The pH and the concentration of aggresive ions in the environment dictate the stability of the passive film. The pitting resistance of materials is determined by electrochemical testing, exposing materials to a chloride-rich environment and low pH \cite{HAMADA2006, TIAN2015}. The associated environmental interactions alter the stability of the passive film and promote pitting \cite{Marcus2002}. This type of corrosion requires consideration in the design as it alters the service life of engineering components operating in aggressive environments. Late detection and prevention can result in premature failure and disastrous events \cite{Uhlig2008}. It is essential to quantify the pH levels and the concentration of aggressive ions to understand the pitting corrosion of materials.

Assessing pitting corrosion damage and measuring the concentration of aggressive ions and pH at the metal$-$electrolyte interface remains challenging due to the unavoidable localized behavior, length, and time scales of the problem. Computational simulations can expand the reach of experimental studies to length and time scales that are difficult to investigate otherwise. Various computational models have been developed to predict the corrosion performance of materials, including Lagrangian-Eulerian approaches \cite{SARKAR2012, SUN2014, Brewick2017}, peridynamics \cite{Jafarzadeh2018, CHEN2015}, level set methods \cite{Duddu2014, Vagbharathi2014}, cellular automata models \cite{WANG2013, FATOBA2018, GONG2022}, and phase-field method \cite{MAI2016, Mai2018, Nguyen2017, Ansari2018, Ansari2019, GAO2020, LIN2021, CUI2021, Tantratian2022, KOVACEVIC2023, CUI2023}. In addition to uniform and galvanic corrosion, pit growth in different materials exposed to various environments is captured with these models. Moreover, the role of mechanical fields \cite{WANG2013, FATOBA2018, GONG2022, CUI2021, Tantratian2022, KOVACEVIC2023}, electrochemistry \cite{Ansari2019,  LIN2021, CUI2023}, and crystallographic orientation \cite{Nguyen2017, Brewick2022, Sahu2022} in governing pit growth, pit-to-crack transition, and crack propagation have been integrated into computational frameworks. However, the motion of ionic species in the electrolyte is simplified and not thoroughly captured in these models. Although surface processes are critical in ion transport, they are neglected in obtaining the solution potential distribution \cite{Ansari2019, Tsuyuki2018, LIN2019, LIN2021, Tantratian2022, Chadwick2018, CUI2023, LIN2020, Chen2022}. More insights into surface effects and mechanisms governing interfacial ion transport remain to be included in computational models.

The formation and charging kinetics of an electric double layer (EDL) formed at the interface between the electrode and the electrolyte play a significant role in driving the transport of ionic species and electric potential \cite{Schmickler2020}. However, despite its contribution to the electrochemical response of the environment, the role of the EDL has not been considered in numerical models \cite{SARKAR2012, SUN2014, Brewick2017, Jafarzadeh2018, CHEN2015, Duddu2014, Vagbharathi2014, WANG2013, FATOBA2018, GONG2022, MAI2016, Mai2018, Ansari2018, Tsuyuki2018, Nguyen2017, Chadwick2018, Ansari2019, GAO2020, LIN2019, LIN2020, LIN2021, CUI2021, Tantratian2022, KOVACEVIC2023, CUI2023, Brewick2022, Sahu2022, Chen2022}. The main obstacle in incorporating the EDL into computational frameworks is ascribed to the nanoscale at which the EDL occurs. The electrochemical phenomena associated with the charging kinetics of the EDL have been considered at a much lower scale \cite{Guyer2004, Guyer2004a, Sherman2017}, which is unpractical for engineering applications. Recently, a phase-field formulation for assessing corrosion in polycrystalline materials that considers the charging kinetics of the EDL has been developed \cite{Makuch2024}. The formulation integrates the effects of the EDL on the motion of ionic species and electric potential without explicitly introducing it in the numerical framework, overcoming the length scale limitation. It utilizes an equivalent circuit model (ECM) to characterize the properties of the EDL, such as its resistance and capacitance. The present work expands on this phase-field formulation \cite{Makuch2024} by examining the role of the properties of the EDL in the motion of species in the solution, electric potential distribution, and corrosion damage. The present investigation provides a more quantitative assessment of corrosion damage, concentration of aggressive ions, and pH within the electrolyte.

The outline of the paper is as follows. The corrosion mechanism and electrochemistry are described in Section \ref{sec2} and the computational framework based on the phase-field method is subsequently formulated. The model is validated against the classic benchmark pencil electrode test \cite{ERNST2002a} in Section \ref{sec3}. After validation, the effect of the properties of the EDL on pit kinetics, current density, and the distribution of electric potential and ionic species are presented. The advantages and disadvantages of the present model, along with the comparison with existing models in the literature, are discussed in Section \ref{sec4}. Conclusions of the investigation are summarised in Section \ref{sec5}.

\section{Computational framework} \label{sec2}
\subsection{Underlying electrochemistry} \label{sec21}
Wet air or aqueous environments break down the protective film formed on metal surfaces, exposing the metal surface to a corrosive environment and initiating corrosion. The following reactions can be used to summarize the corrosion process \cite{Uhlig2008}
\begin{equation} \label{eqn1}
    \text{M}_{(s)} \rightarrow \text{M}_{(aq)}^{z_1+} + z_1e^- \text{ (metal dissolution)}
\end{equation}
\begin{equation} \label{eqn2}
    \text{M}_{(aq)}^{z_1+} + \text{H}_2\text{O} \xrightleftharpoons[k_{1b}]{k_{1f}} \text{M}(\text{OH})^{(z_1-1)+} + \text{H}^+ \text{ (primary hydrolysis)}
\end{equation}
\begin{equation} \label{eqn3}
    \text{H}_2\text{O} \xrightleftharpoons[k_{2b}]{k_{2f}} \text{OH}^- + \text{H}^+ \text{ (water dissociation)},
\end{equation}
where M is the corroded metal, $z_1$ is the charge number, $k_{1f}$, $k_{2f}$, $k_{1b}$, and $k_{2b}$ are the forward and the backward reaction constants. The description in equations (\ref{eqn1}), (\ref{eqn2}) and (\ref{eqn3}) does not consider the precipitation of stable phases such as salts and oxides. The oxides are not treated in the present model under the assumption that the simulated environment is harsh enough to prevent surface passivation. The presence of salts is incorporated as the equilibrium concentration of metal ions in the liquid phase, as discussed below in Section \ref{sec22} and Section \ref{sec31}.

\subsection{Kinetics and thermodynamics} \label{sec22}

\begin{figure}[h!]
    \centering
    \includegraphics[width = 15 cm]{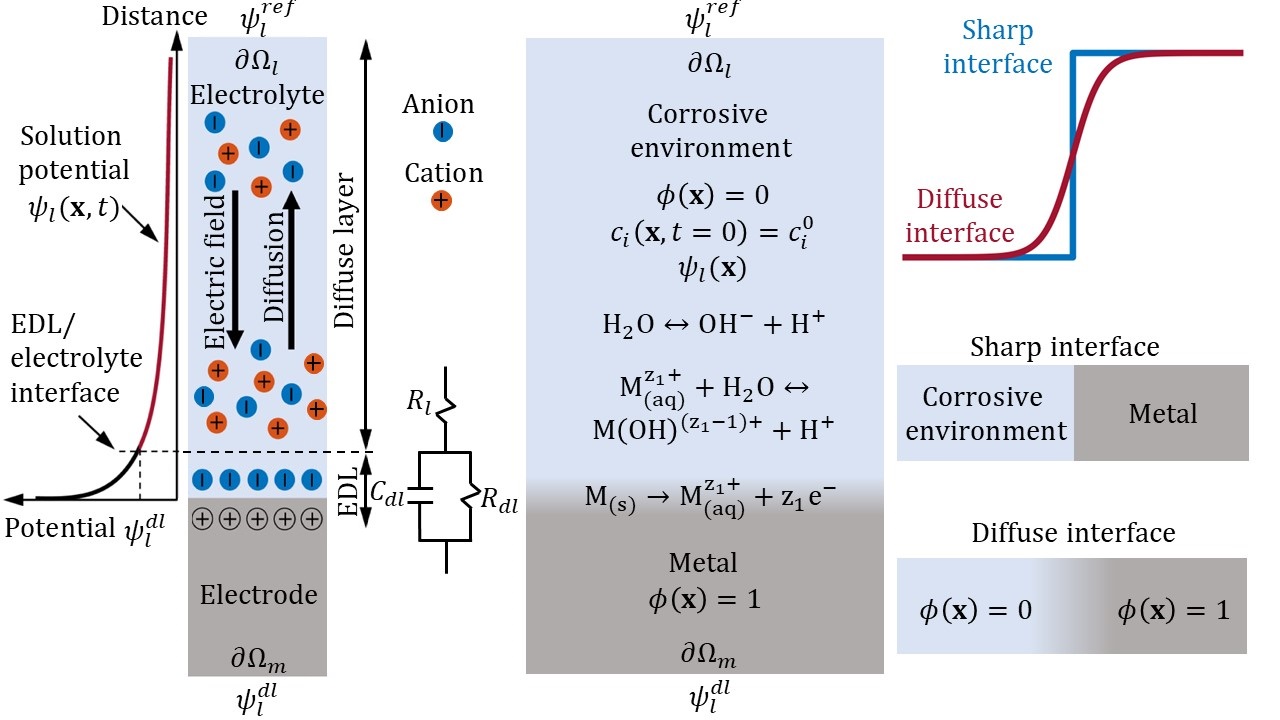}
    \captionsetup{labelfont = bf,justification = raggedright}
    \caption{Schematic of equivalent circuit model and phase-field description of the metal and corrosive environment phases.}
    \label{Fig1}
\end{figure}

Figure \ref{Fig1} depicts the electrochemical system considered. A set of concentrations of ionic species represented by $\overrightarrow{c} = (c_1 = \text{M}^{z_1+}, c_2 = \text{M(OH)}^{(z_1-1)+}, c_3 = \text{H}^+, c_4 = \text{OH}^-, c_5 = \text{Na}^+, c_6 = \text{Cl}^-)$ describes the reactions in equations (\ref{eqn1}), (\ref{eqn2}) and (\ref{eqn3}) and the surrounding environment. Although Na$^+$ and Cl$^-$ ions are not considered in the above reactions, their presence influences the movement and distribution of remaining ions since they are charged particles. The conductivity of the electrolyte is a function of all ionic species and influences the solution potential distribution (Section \ref{sec233}). The electrode and the electrolyte domains are distinguished by the phase-field variable: $\phi = 1$ represents the electrode, $\phi = 0$ corresponds to the electrolyte, and $0 < \phi < 1$ indicates the thin interfacial region between the phases (electrode$-$electrolyte interface). The independent kinematic variables necessary for the model description are the phase-field parameter that describes the evolution of the corroding interface $\phi(\mathbf{x},t)$, the concentration variable $c_{i}(\mathbf{x},t)$ for each ionic species considered, and the electric potential $\psi_l(\mathbf{x},t)$.

The free energy functional of a heterogeneous system occupying the domain $\Omega$ (figure \ref{Fig1}) with contributions from the chemical free energy density $f^{chem}$, interfacial energy density $f^{grad}$, and electric free energy density $f^{elec}$ can be represented as
\begin{equation} \label{eqn4}
                   \mathscr{F} = \int_\Omega \Big[f^{chem}(\overrightarrow{c},\phi) + f^{grad}(\nabla\phi) + f^{elec}(\overrightarrow{c},\psi_l)\Big]\,d\Omega.
\end{equation}

The chemical free energy density $f^{chem}$ is given as 
\begin{equation} \label{eqn5}
                   f^{chem}(\overrightarrow{c},\phi) = f^{chem}_1(\bar{c}_1, \phi) + \sum_{i = 2} f^{chem}_{i}(\bar{c}_i) + \omega g(\phi),
\end{equation}
where $f^{chem}_1(\bar{c}_1, \phi)$ and $f^{chem}_{i}(\bar{c}_i)$ ($i\neq1$) are the chemical free energy densities as a function of normalized ion concentrations $\bar{c}_i = c_i V_m$. $V_m$ denotes the molar volume of the metal. In the previous equation, $g(\phi) = 16\phi^2(1-\phi)^2$ is the double-well potential with minima at $\phi = 0$ (electrolyte) and $\phi = 1$ (electrode) and $\omega = 3\Gamma/(4\ell)$ represents the energy barrier height at $\phi = 1/2$ \cite{KOVACEVIC2020}. Here, $\Gamma$ is the interfacial energy and $\ell$ is the chosen interface thickness. The chemical free energy density $f^{chem}_1(\bar{c}_1,\phi)$ is represented as \cite{KKS1999}
\begin{equation} \label{eqn6}
             f^{chem}_1(\bar{c}_1,\phi) = (1-h(\phi))f^{chem}_{l}(\bar{c}_1^l) + h(\phi)f^{chem}_{s}(\bar{c}_1^s),
\end{equation}
where $f^{chem}_{l}(\bar{c}_1^l)$ and $f^{chem}_{s}(\bar{c}_1^s)$ stand for the chemical free energy densities of the liquid and solid phases. They are expressed as a function of normalized metal ion phase-concentrations within the liquid $\bar{c}_1^l$ and the solid phases $\bar{c}_1^s$. $h(\phi) = \phi^3(10-15\phi+6\phi^2)$ is a monotonically increasing interpolation function. For simplicity, the chemical free energy densities $f^{chem}_{l}(\bar{c}_1^l)$ and $f^{chem}_{s}(\bar{c}_1^s)$ are represented by parabolic functions with the same density curvature parameter $A$ as
\begin{equation} \label{eqn7}
                   f^{chem}_{l}(\bar{c}_1^l) = \frac{1}{2}A(\bar{c}_1^l - \bar{c}_1^{l,eq})^2  \quad\mathrm{}\quad  f^{chem}_{s}(\bar{c}_1^s) = \frac{1}{2}A(\bar{c}_1^s - \bar{c}_1^{s,eq})^2,
\end{equation}
where $\bar{c}_1^{l,eq} = c_1^{l,eq} V_m$ and $\bar{c}_1^{s,eq} = c_1^{s,eq} V_m = 1$ are the normalized equilibrium phase concentrations. Alternative ways for describing the chemical free energy density using first-principles calculations and thermodynamic databases \cite{KUMARTHAKUR2023} will be addressed in future work. The parameter $A$ serves as an energetic penalty for departing from the equilibrium concentration \cite{KOVACEVIC2022}. The equilibrium phase-concentration in the liquid phase $c_1^{l,eq}$ is determined based on the solubility of salts formed on the exposed metal surface as the formation of corrosion products and the passive film are neglected in the present corrosion mechanism, equations (\ref{eqn1}), (\ref{eqn2}), and (\ref{eqn3}). The normalized concentration of metal ions is given as a function of $\bar{c}_1^l$ and $\bar{c}_1^s$: $\bar{c}_1 = (1-h(\phi)) \bar{c}_1^l + h(\phi)\bar{c}_1^s$. It is further assumed in the present model that the same diffusion chemical potential ($\partial f^{chem}_{l}(\bar{c}_1^l) / \partial \bar{c}_1^l = \partial f^{chem}_{s}(\bar{c}_1^s) / \partial \bar{c}_1^s$) holds within the interfacial region \cite{KKS1999}. Built upon this assumption, the chemical free energy density $f^{chem}_1(\bar{c}_1,\phi)$ is written as \cite{Makuch2024}
\begin{equation} \label{eqn8}
    f^{chem}_1(\bar{c}_1,\phi) = \frac{1}{2} A \Big[\bar{c}_1 -h(\phi) (\bar{c}_1^{s,eq} - \bar{c}_1^{l,eq}) - \bar{c}_1^{l,eq}\Big]^2.
\end{equation}
The dilute solution theory is used to express the contribution to the chemical free energy density from the other ions present in the electrolyte \cite{Bazant2013}
\begin{equation} \label{eqn9}
                   f^{chem}(\bar{c}_i) = \frac{RT}{V_m}\sum_{i = 2} \bar{c}_i\ln \bar{c}_i + \sum_{i = 2} \bar{c}_i \mu_{i}^{\Theta},
\end{equation}
where $\mu_{i}^{\Theta}$ is the reference chemical potential of ionic species $i$, $R$ is the universal gas constant, and $T$ is the absolute temperature.

The interfacial free energy density is defined as
\begin{equation} \label{eqn10}
                   f^{grad}(\nabla\phi) =\frac{1}{2}\kappa|\nabla \phi|^2,
\end{equation}
where $\kappa = 3\Gamma\ell/2$ is the isotropic gradient energy coefficient \cite{KOVACEVIC2020}.

The electric free energy density of a charged assembly subjected to a net solution potential is given as \cite{Jefimenko1989}
\begin{equation}\label{eqn11}
    f^{elec}(\overrightarrow{c},\psi_l) =  \psi_l F\sum_{i}z_i c_i,
\end{equation}
where $F$ is Faraday's constant and $z_i$ is the charge number of component $i$. 

\subsection{Governing equations} \label{sec23}
\subsubsection{Interface evolution and electro-chemical coupling}

The evolution of the metal$-$electrolyte interface follows the Allen-Cahn equation \cite{ALLEN19791085}
\begin{equation} \label{eqn12}
\frac{\partial \phi}{\partial t} = - L\frac{\delta \mathscr{F}}{\delta \phi} = -L\Big(\frac{\partial f^{chem}}{\partial\phi} - \kappa\nabla^2\phi \Big)  \quad \text{in}\quad\Omega; \quad\mathrm{} \kappa \mathbf{n} \cdot \nabla \phi = 0 \quad \text{on}\quad\partial\Omega,
\end{equation}
where $L > 0$ is the phase-field mobility parameter. In the present model, the definition of phase-field mobility $L$ is enhanced to account for the role of local current density in controlling the motion of the electrode$-$electrolyte interface. The local current density is written as \cite{Jones1996}
\begin{equation} \label{eqn13}
\begin{aligned}
i_a= i_0 \Big[\text{exp}\Big (\frac{\alpha z_1 F \eta}{RT}\Big) - \text{exp}\Big(-\frac{(1 - \alpha) z_1 F \eta}{RT}\Big)\Big] \quad\mathrm{}\quad \eta = E_{app} - E_{eq}-\psi_{l},
\end{aligned}
\end{equation}
where $i_0$ is the exchange current density, $\alpha$ is the anodic charge transfer coefficient ($\alpha = 0.26$ in this work \cite{CUI2023}), $\eta$ is the overpotential, $E_{app}$ is the applied electric potential, and $E_{eq}$ is the equilibrium corrosion potential. Using the previous equation and the direct proportionality between the phase-field mobility and the current density \cite{MAI2016, CUI2022}, an enhanced definition of interfacial mobility can be obtained
\begin{equation}\label{eqn14}
        L = L_0 \Big[\text{exp}\Big (\frac{\alpha z_1 F \eta}{RT}\Big) - \text{exp}\Big(-\frac{(1 - \alpha) z_1 F \eta}{RT}\Big)\Big],
\end{equation}
where $L_0$ is the interfacial mobility that corresponds to the exchange current density $i_0$. The mobility parameter $L_0$ is calibrated against experimental measurements in Section \ref{sec3}. The motion of the interface is nonlinearly proportional to the overpotential following equation (\ref{eqn13}). It is demonstrated in Appendix A that this enhanced definition of interfacial mobility captures a nonlinear trend in current density and interface velocity with respect to overpotential. The present model with this definition of interfacial mobility resembles nonlinear phase-field models formulated on reaction rate theory \cite{Linyun2012, Liang2014, CHEN2015a}.

\subsubsection{Mass transport}

The mass transport of ionic species considered is written as \cite{Makuch2024}
\begin{equation} \label{eqn15}
\left\{
\begin{aligned}
& \frac{\partial c_1 }{\partial t} = -\nabla \cdot \mathbf{J}_1 + R_1; \quad\mathrm{}\quad \mathbf{J}_1 = -D_1\Big(\nabla c_1 - h^{\prime}(\phi)(c_1^{s,eq} - c_1^{l,eq}) \nabla \phi + \frac{z_1 F}{R T} c_1 \nabla \psi_l \Big)\\
& \frac{\partial c_i}{\partial t} = -\nabla \cdot \mathbf{J}_i + R_i; \quad\mathrm{}\quad \mathbf{J}_i = -D_i \Big(\nabla c_i + \frac{z_i F}{R T} c_i \nabla \psi_l \Big) \quad\mathrm{}\quad i \neq 1\\
\end{aligned}
\right\},\
\end{equation}
where $\mathbf{J}_i$ denotes the electrochemical flux and $R_i$ are the volumetric reaction rates associated with the reactions given in equations (\ref{eqn2}) and (\ref{eqn3}). On the boundary $\partial\Omega$: $\mathbf{n}\cdot \mathbf{J}_i = 0$. The dilute solution theory is used to derive the electromigration flux in the equation for the metal ion (the last term in $\mathbf{J}_1$) \cite{Ansari2018, Mai2018, CUI2023}. $D_i$ stands for the effective diffusion coefficient given as: $D_{i} = D^s_{i}h(\phi)+(1-h(\phi))D^l_{i}$, where $D^l_{i}$ and $D^s_{i}$ are the diffusion coefficients of ions in the electrolyte and metal phases ($D^s_{i} \ll D^l_{l}$). The volumetric chemical reaction rates $R_i$ are \cite{Makuch2024}
\begin{equation} \label{eqn16}
\begin{aligned}
R_1 = k_{1b}(c_2 c_3-K_1 c_1) \quad\mathrm{}\quad R_2 = -R_1 \quad\mathrm{}\quad R_3 = R_2 + R_4 \quad\mathrm{}\quad R_4 = k_{2b}(K_2 - c_3 c_4),
\end{aligned}
\end{equation}
where $K_1 = k_{1f}/k_{1b} = c_2 c_3/c_1$ and $K_2 = k_{2f}/k_{2b} = c_3 c_4$ are the equilibrium constants for the chemical reaction in equations (\ref{eqn2}) and (\ref{eqn3}) \cite{Baes1976}. $k_{1b}$ and $k_{2b}$ are the penalty coefficients that enforce the equilibrium in the electrolyte. $R_1$, $R_2$, $R_3$, and $R_4$ are the volumetric reaction rates ascribed to M$^{z_{1}+}$, M(OH)$^{(z_1 -1)+}$, H$^+$, and OH$^{-}$ ions. $R_5 = R_6 = 0$ as Na$^+$ and Cl$^-$ ions are not considered in equations (\ref{eqn1}), (\ref{eqn2}), and (\ref{eqn3}).

\subsubsection{Solution potential}\label{sec233}

The interface between the electrode and the electrolyte spontaneously acquires a charge that forms an electric double layer (EDL), figure \ref{Fig1}. These surface charges prevent the motion of ions within the EDL. They play a critical role in the distribution and motion of ions beyond this layer \cite{Schmickler2020}. The solution potential distribution outside the EDL can be obtained from the following equation \cite{Jefimenko1989} 
\begin{equation}\label{eqn17}
\begin{aligned}
    & \nabla \cdot \lambda \nabla \psi_l = 0 \quad\mathrm{} \text{in}\quad\Omega \\
    \quad\mathrm{} \lambda \mathbf{n}\cdot \nabla \psi_l = 0 \quad\text{on}\quad\partial\Omega \quad &\psi_l = \psi_l^{dl} \quad\text{on}\quad\partial\Omega_m \quad\text{and} \quad \psi_l = \psi_l^{ref} \quad\text{on}\quad\partial\Omega_l,
\end{aligned}
\end{equation}
where $\psi_l^{ref}$ is the solution potential on the reference electrode boundary $\partial\Omega_l$ and $\psi_l^{dl}$ is the potential at the interface between the EDL and the diffuse layer, figure \ref{Fig1}. In equation (\ref{eqn17}), $\lambda = \lambda^s h(\phi) + (1-h(\phi)) \lambda^l$ is the effective conductivity expressed as a function of conductivities in the liquid $\lambda^l$ and solid phases $\lambda^s$ ($\lambda^s \gg \lambda^l$). The electric conductivity in the electrolyte is given as a function of ionic species concentrations: $\lambda_{l} = \sum_{i} D_{i}F^{2}z_{i}^{2}c_{i}/(RTV_{m})$ \cite{Neueder2014}. A first-order resistor-capacitor ECM, given in figure \ref{Fig1}, is used to determine the evolution of the potential at the EDL$-$electrolyte interface $\psi_l^{dl}$
\begin{equation}\label{eqn18}
\psi_l^{dl} = \psi^0 \Big(\frac{1}{1 + \chi} + \frac{\chi}{1 + \chi} \text{exp} \big( -\frac{t \text{ln} 2}{\xi t_c} \big) \Big),
\end{equation}
where $\psi^0$ is the initial surface polarization at time zero ($t=0$), $\chi$ is the proportionality constant between the EDL resistance $R_{dl}$ and the solution resistance $R_{dl}$ (figure \ref{Fig1}), $\xi$ is the geometric factor that accounts for the change in the electrode$-$electrolyte interfacial area during the dissolution process, and $t_{c}$ is the half-time of capacitor charging associated with the formation of the capacitance of the EDL \cite{Sundararajan2020}. The magnitude of $\psi_l^{dl}$ obtained using equation (\ref{eqn18}) is enforced on the metal boundary $\partial\Omega_m$, figure \ref{Fig1}. The high conductivity of the metal $\lambda^s$ ensures that there is no electric potential drop within the metal phase. This ensures that the electric potential at the metal$-$electrolyte interface is the same as the value described by equation (\ref{eqn18}). In that way, the effect of the EDL on the evolution of the solution potential is accounted for without explicitly introducing it in the computational domain. The ECM parameters $\chi$ and $t_c$ in equation (\ref{eqn18}) can be calibrated against experimental measurements of current density or extracted from experimental studies, as illustrated in Section \ref{sec31}. The effect of $\chi$ and $t_c$ on pit kinetics, current density, and environmental response in terms of the distribution of ionic species and solution potential is shown in the sensitivity analyses in Section \ref{sec32} and Section \ref{sec33}. Appendix B provides the details regarding the derivation of equation (\ref{eqn18}) based on the first-order resistor-capacitor ECM used in the present work. A similar equation for $\psi_l^{dl}$ utilizing another frequently employed circuit model in experimental practice is also provided in Appendix B.   

The resulting set of governing equations includes equations (\ref{eqn12}), (\ref{eqn15}), and (\ref{eqn17}) along with the accompanying boundary conditions. The details of the numerical implementation of the governing equations, finite element mesh, solver tolerance and solution algorithm, and mesh sensitivity analysis are given in \cite{Makuch2024}. The code developed is available at \url{https://mechmat.web.ox.ac.uk/codes}.

\section{Results} \label{sec3}

\subsection{Model calibration and validation} \label{sec31}

\begin{figure}[h!]
    \centering
    \includegraphics[width = 12cm]{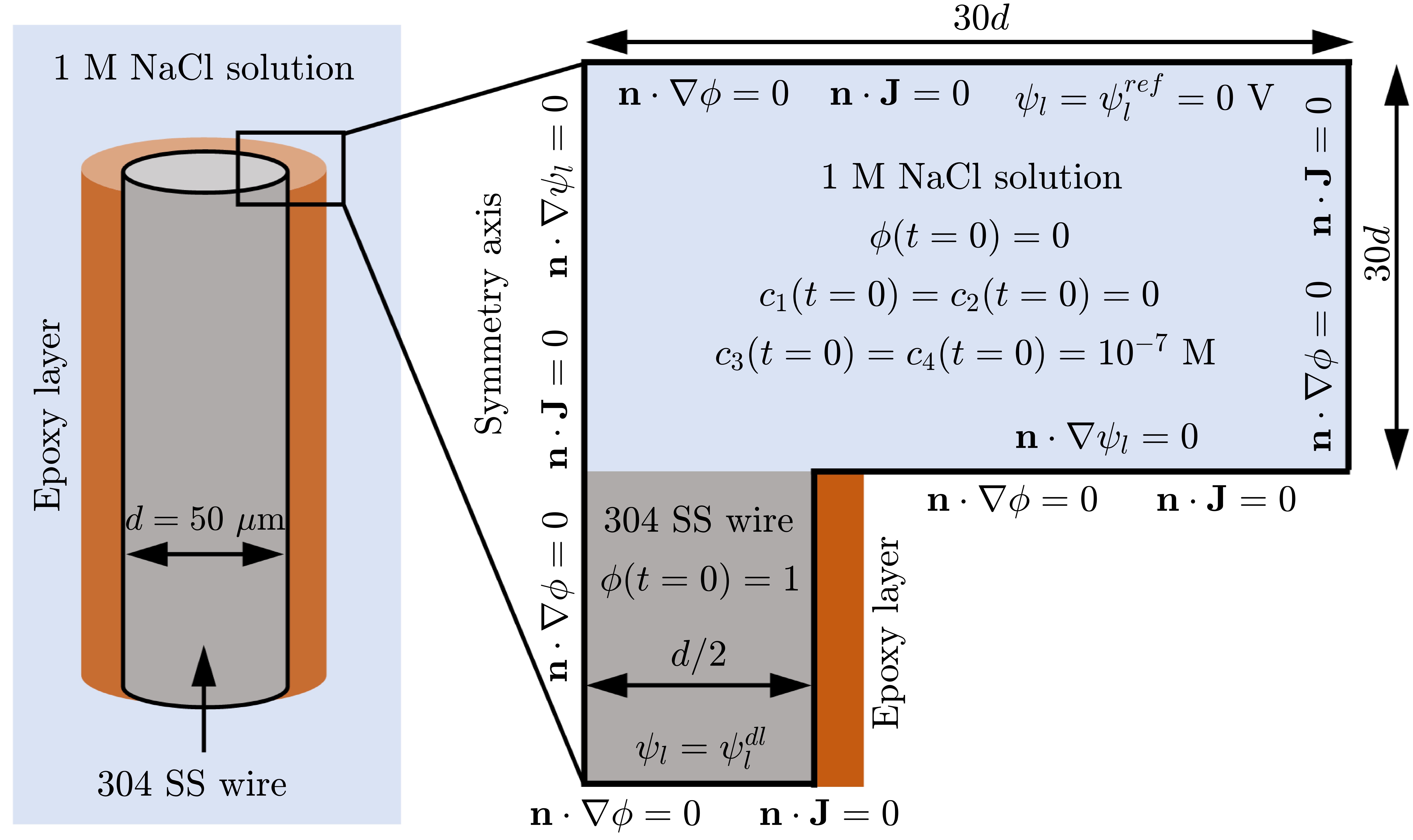}
    \captionsetup{labelfont = bf,justification = raggedright}
    \caption{Illustration of the experimental setup used in Ref. \cite{ERNST2002a} (left) and the corresponding computational domain (right).}
    \label{Fig2}
\end{figure}

The classic benchmark pencil electrode test is used to validate the computational framework developed, figure \ref{Fig2}. The experiment consisted of a 304 stainless steel (SS) wire with a 50 $\mu$m diameter immersed in 1 M NaCl solution \cite{ERNST2002a}. An epoxy resin was used to coat the wire circumferentially. The coating prevented dissolution in the radial direction, leaving only the cross-section surface exposed to the corrosive environment. The potential difference between the working and the reference saturated calomel electrode (SCE) was 600 mV. Experimental measurements of pit depth and current density were recorded during the test. The experimental data is used to calibrate the interface mobility parameter $L_0$ and the ECM parameters $\chi$ and $t_c$.

An axisymmetric domain depicted in figure \ref{Fig2} is considered in the phase-field simulation. The diameter of the metal wire is the same as in the experiment, $d$ = 50 $\mu$m. The size of the liquid phase is selected to be much larger than the metal to mimic the experimental setup \cite{ERNST2002a}. The large computational domain prevents the electrolyte from being saturated with metal ions. The resulting set of the governing equations is solved with accompanying initial and boundary conditions. A smooth phase-field profile is prescribed as the initial electrode$-$electrolyte interface to initiate the solid ($\phi = 1$) and liquid ($\phi = 0$) phases. The initial concentration of metal ions M$^{z_{1}+}$ in the solid phase is defined using the molar volume of the metal $V_m$. The initial concentrations of metal ions M$^{z_{1}+}$ and metal hydroxide ions M(OH)$^{(z_1-1)+}$ in the liquid phase are equal to zero. The other ionic species considered are only present in the electrolyte. A neutral pH of 7 is employed to determine the initial concentrations of H$^{+}$ and OH$^{-}$ ions in the electrolyte. The solution salinity used in the experiment, figure \ref{Fig2}, returns the initial concentration of Na$^{+}$ and Cl$^{-}$ ions in the electrolyte domain. The initial electric potential distribution is set to 0 V (vs. SCE). No flux boundary condition for all kinematic variables is applied to all domain boundaries except on the top and bottom boundaries. A reference solution potential of 0 V (vs. SCE) is prescribed on the far top boundary in the electrolyte. The electric potential at the interface between the EDL and the electrolyte $\psi_l^{dl}$ (see equation (\ref{eqn18})) is enforced on the bottom metal surface, figure \ref{Fig2}. The narrow electrode size and epoxy coating enforce no change in the electrode$–$electrolyte interfacial area, and thus, the geometric factor in equation (\ref{eqn18}) is set to $\xi$ = 1.

The parameters used in all simulations in this study are listed in Table \ref{table1}. The penalty constants $k_{1b}$ and $k_{2b}$ are selected to be large enough to enforce the equilibrium conditions in equation (\ref{eqn16}) but not so large as to hinder numerical convergence. The parametric sensitivity analysis for different values of the penalty constants is conducted in Appendix C. The equilibrium corrosion potential $E_{eq}$, the charge number $z_1$, and the concentration $c_{1}^{s,eq}$ are defined using the molar fraction of alloying elements. The charge numbers of the other ions are given in equations (\ref{eqn2}) and (\ref{eqn3}). The equilibrium concentration in the liquid phase $c_{1}^{l,eq}$ is obtained using the solubility of salts formed on the exposed metal surface in this metal$-$environment system \cite{Isaacs1995}. The formation of salts is not considered in the present model. The chemical free energy density curvature parameter $A$ is assumed from similar phase-field studies \cite{MAI2016, CUI2021, KOVACEVIC2023}. 

\begin{table} [h!]
\centering
\captionsetup{labelfont = bf,justification = raggedright}
\caption{Parameters used in all simulations in this study.}
\begin{tabular}{l l l} 
\hline
Quantity & Value & Unit\\
\hline
Equilibrium concentration in the solid phase $c_1^{s,eq}=1/V_m$ & 144.3 & mol/L \cite{BS2005}\\   
Equilibrium concentration in the liquid phase $c_1^{l,eq}$ & 5.1 & mol/L \cite{Isaacs1995}\\
Diffusion coefficient of M$^{z_1+}$ in the liquid phase $D^l_{1}$ & $0.719 \times 10^{-9}$ & m/s$^2$ \cite{Haynes2016}\\
Diffusion coefficient of M(OH)$^{(z_1-1)+}$ in the liquid phase $D^l_{2}$ & $0.719 \times 10^{-9}$ & m/s$^2$ \cite{Haynes2016}\\
Diffusion coefficient of H$^{+}$ in the liquid phase $D^l_{3}$ & $9.311 \times 10^{-9}$ & m/s$^2$ \cite{Haynes2016}\\
Diffusion coefficient of OH$^{-}$ in the liquid phase $D^l_{4}$ & $5.273 \times 10^{-9}$ & m/s$^2$ \cite{Haynes2016}\\
Diffusion coefficient of Na$^{+}$ in the liquid phase $D^l_{5}$ & $1.334 \times 10^{-9}$ & m/s$^2$ \cite{Haynes2016}\\
Diffusion coefficient of Cl$^{-}$ in the liquid phase $D^l_{6}$ & $2.032 \times 10^{-9}$ & m/s$^2$ \cite{Haynes2016}\\
Interfacial energy $\Gamma$ & 2.10 & J/m$^2$ \cite{Kanhaiya2014}\\
Interface thickness $\ell$ & 5 & $\mu$m\\
Chemical free energy density curvature parameter $A$ & $1.02\times10^8$ & J/m$^3$ \cite{CUI2021, KOVACEVIC2023}\\
Primary hydrolysis equilibrium constant $K_{1}$ & $3.1622 \times 10^{-7}$ & mol/m$^3$ \cite{Baes1976}\\
Water dissociation equilibrium constant $K_{2}$ & $10^{-8}$ & mol$^2$/m$^6$ \cite{Baes1976}\\
Equilibrium corrosion potential $E_{eq}$ & -0.729 & V (vs. SCE) \cite{BS2005}\\
Solid phase conductivity $\lambda_{s}$ & 10$^6$ & S/m \cite{BS2005}\\
\hline
\end{tabular}
\label{table1}
\end{table}

\begin{figure}[h!]
    \centering
    \includegraphics[width = 15cm]{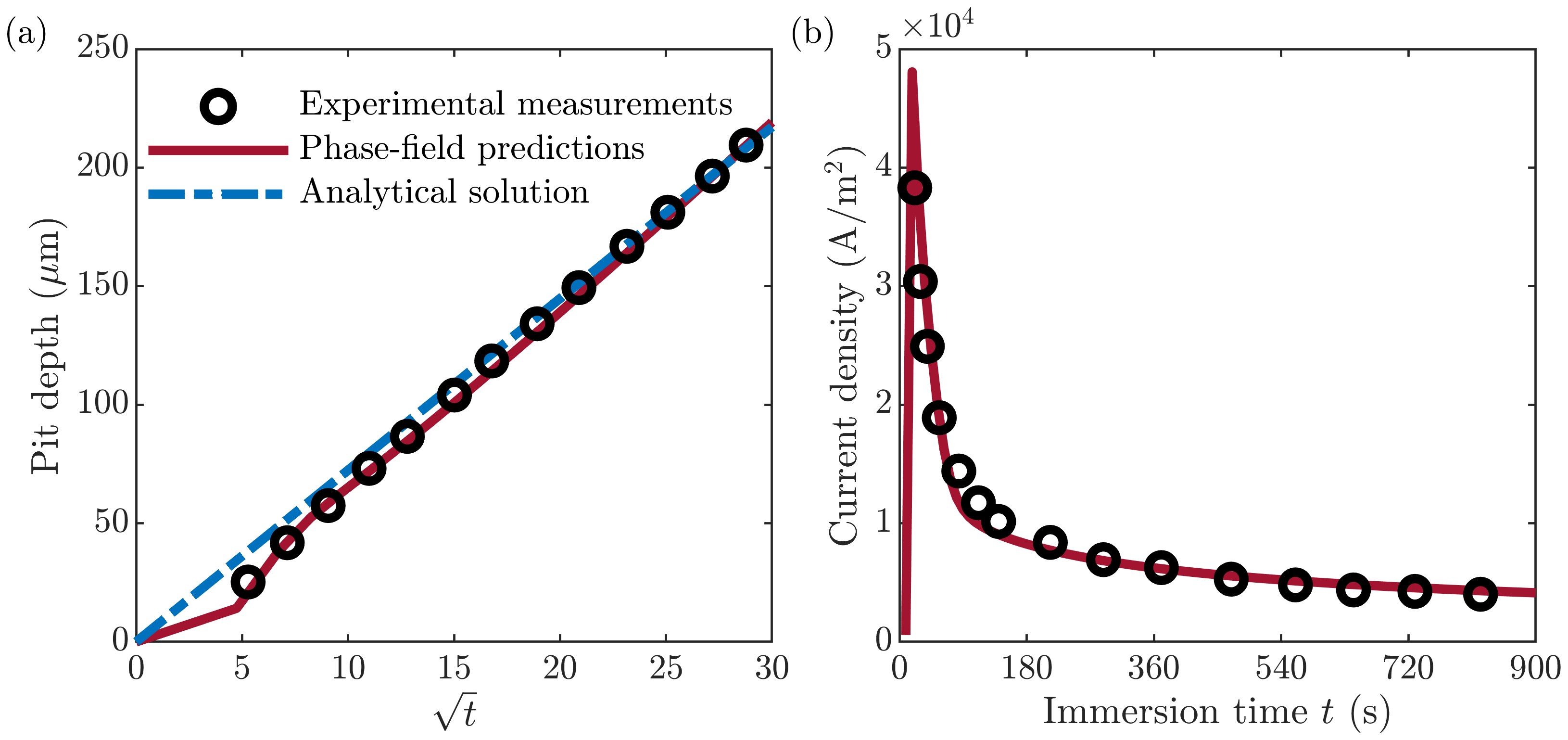}
    \captionsetup{labelfont = bf,justification = raggedright}
    \caption{Comparison between experimental measurements \cite{ERNST2002a}, analytical solution \cite{SCHEINER2007}, and phase-field predictions of (a) the evolution of pit depth and (b) current density as a function of immersion time.}
    \label{Fig3}
\end{figure}

Comparison between the model predictions and experimental results \cite{ERNST2002a} in terms of pit depth and current density is given in figure \ref{Fig3}. The model predictions are in good agreement with the experimental data. An analytical solution of pit depth evolution can be derived considering an equivalent 1D problem \cite{SCHEINER2007}. The analytical solution assumes that the transport of ions far from the interface is the rate-limiting process (diffusion$-$controlled corrosion) and that diffusion is the only contribution to ionic migration. The analytical solution returns a linear relation between pit depth and the square root of the immersion time, figure \ref{Fig3}(a). However, a departure from this linear trend can be observed in the experimental data and model predictions at early immersion times, figure \ref{Fig3}(a). This deviation from the analytical solution indicates that the initial pit kinetics is in an activation-controlled regime. The linear trend in pit kinetics and peak in current density (figure \ref{Fig3}) at early immersion times is associated with the charging of the EDL. The highest magnitude of solution potential is achieved for immersion times comparable to $t_c$, equation (\ref{eqn18}). At these times, the overpotential in equation (\ref{eqn13}) is significantly reduced due to high solution potential. Reduced overpotential leads to a lower kinetic coefficient $L$ (equation \ref{eqn14}), returning slower pit propagation. Once immersion time exceeds $t_c$, the process transitions to diffusion$-$controlled. This change in the rate-limiting process from activation-controlled to diffusion-controlled is ascribed to the electric potential drop at the EDL, which builds up enough to release metal ions. The experimental measurements and phase-field predictions follow the analytical solution in the diffusion-controlled stage. The phase-field predictions in figure \ref{Fig3} are obtained using $L_0 = 1.2\times 10^{-15}$ m$^3$/(J~s), $\chi = 120$, and $t_c = 10$ s. 

\begin{figure}[h!]
    \centering
    \includegraphics[width = 15cm]{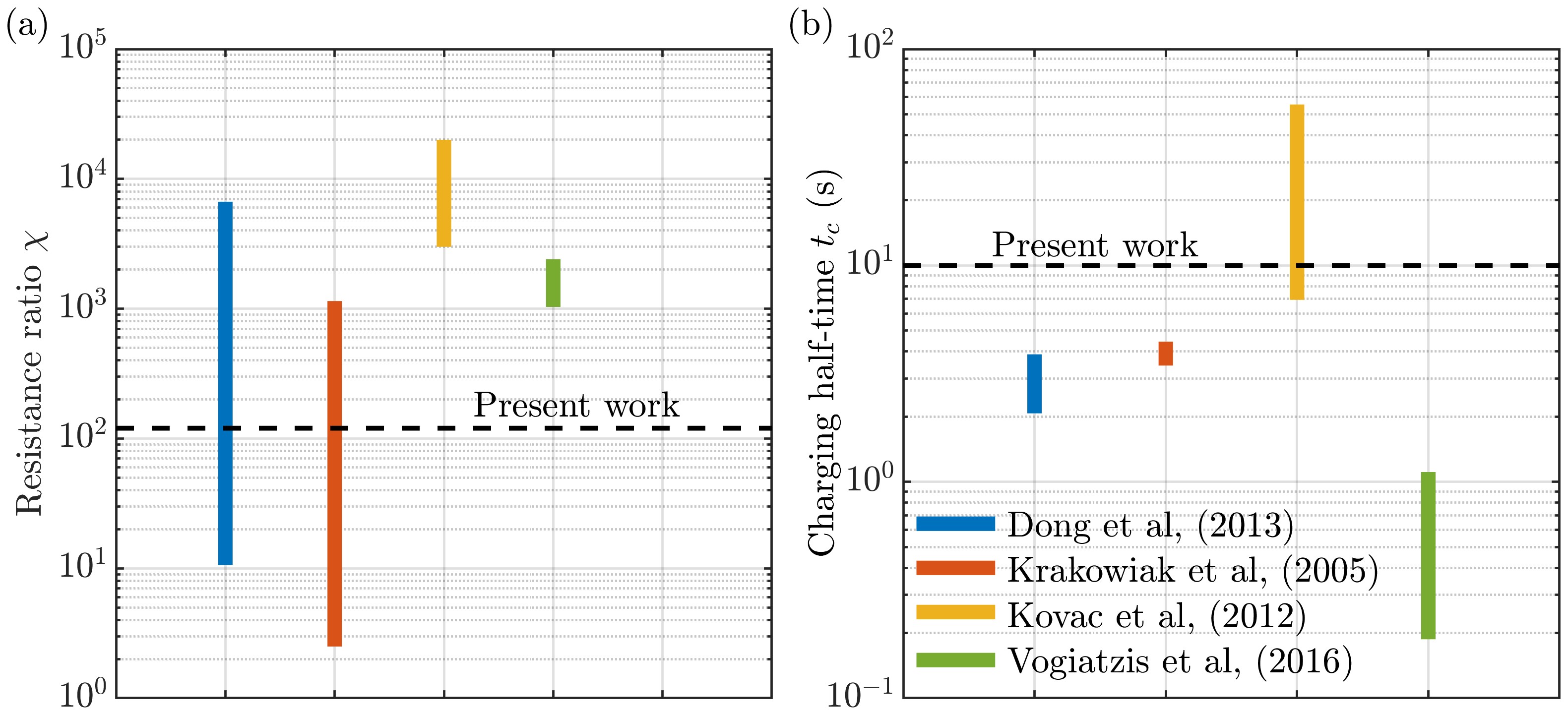}
    \captionsetup{labelfont = bf,justification = raggedright}
    \caption{Equivalent circuit model parameters used in experimental studies (Dong et al. \cite{Dong2013}, Krakowiak et al. \cite{KRAKOWIAK2005}, Kovac et al. \cite{Kovac2012}, Vogiatzis et al. \cite{Vogiatzis2016}) and the present model. (a) Resistance proportionality constant $\chi$ and (b) half-time of capacitor charging $t_c$.}
    \label{Fig4}
\end{figure}

The obtained ECM parameters $\chi$ and $t_c$ are compared with experimental data on 304 SS immersed in a similar aqueous solution \cite{Dong2013, KRAKOWIAK2005, Kovac2012, Vogiatzis2016}. The same equivalent circuit diagram (figure \ref{Fig1}) has been employed in these references to determine the properties of the EDL. The experimentally obtained values for $\chi$ and $t_c$, together with the values used in the present model, are given in figure \ref{Fig4}. The results in figure \ref{Fig4} indicate that the constants $\chi$ and $t_c$ used in the phase-field simulation of the pencil electrode test follow the experimental data. Variations in experimentally obtained values for $\chi$ and $t_c$ can be attributed to different environmental conditions. A sensitivity analysis is conducted in the following section to investigate the effect of ECM parameters on the pit depth and current density response.

\subsection{Sensitivity analysis for equivalent circuit model parameters} \label{sec32}

The sensitivity study is performed for the ECM parameters $\chi$ and $t_c$. Three cases are considered for the resistance ratio ($\chi = 20$, $\chi = 120$ and $\chi = 1200$) and half-time of capacitor charging ($t_c = 1$ s, $t_c = 10$ s and $t_c = 100$ s). The values $\chi = 120$ and $t_c = 10$ s are adopted in the previous section for model calibration. This case study is used for comparison. The same material properties and numerical parameters, given in Section \ref{sec31} and Table \ref{table1}, are used in the sensitivity analysis. Both activation and diffusion-controlled rate-limiting processes are considered in the sensitivity study. Applied potentials of -479 mV and 600 mV (vs. SCE) are employed to trigger the activation and diffusion-controlled processes. The former corresponds to the natural dissolution of metals with an overpotential of 250 mV \cite{Uhlig2008}. The latter is used above for model validation (Section \ref{sec31}).

Interface reactions and short-range interactions dictate the corrosion rate in the case of the small applied potential (-479 mV (vs. SCE)). The time-scale of corrosion damage in such an activation-controlled process is significantly larger than the half-time of capacitor charging (i.e., $t \gg t_c$). Under this condition, the parameter $t_c$ has a negligible effect on pit depth and current density as the second term in equation (\ref{eqn18}) tends to zero for $t \gg t_c$. The process only depends on the resistance ratio $\chi$. The pit kinetics and current density response are identical for all the values of $t_c$ considered. Moreover, the values of $\chi$ considered have a minor effect on the response of the system. Hence, these results for an activation-controlled process are not shown here for brevity.

\begin{figure}[h!]
    \centering
    \includegraphics[width = 15cm]{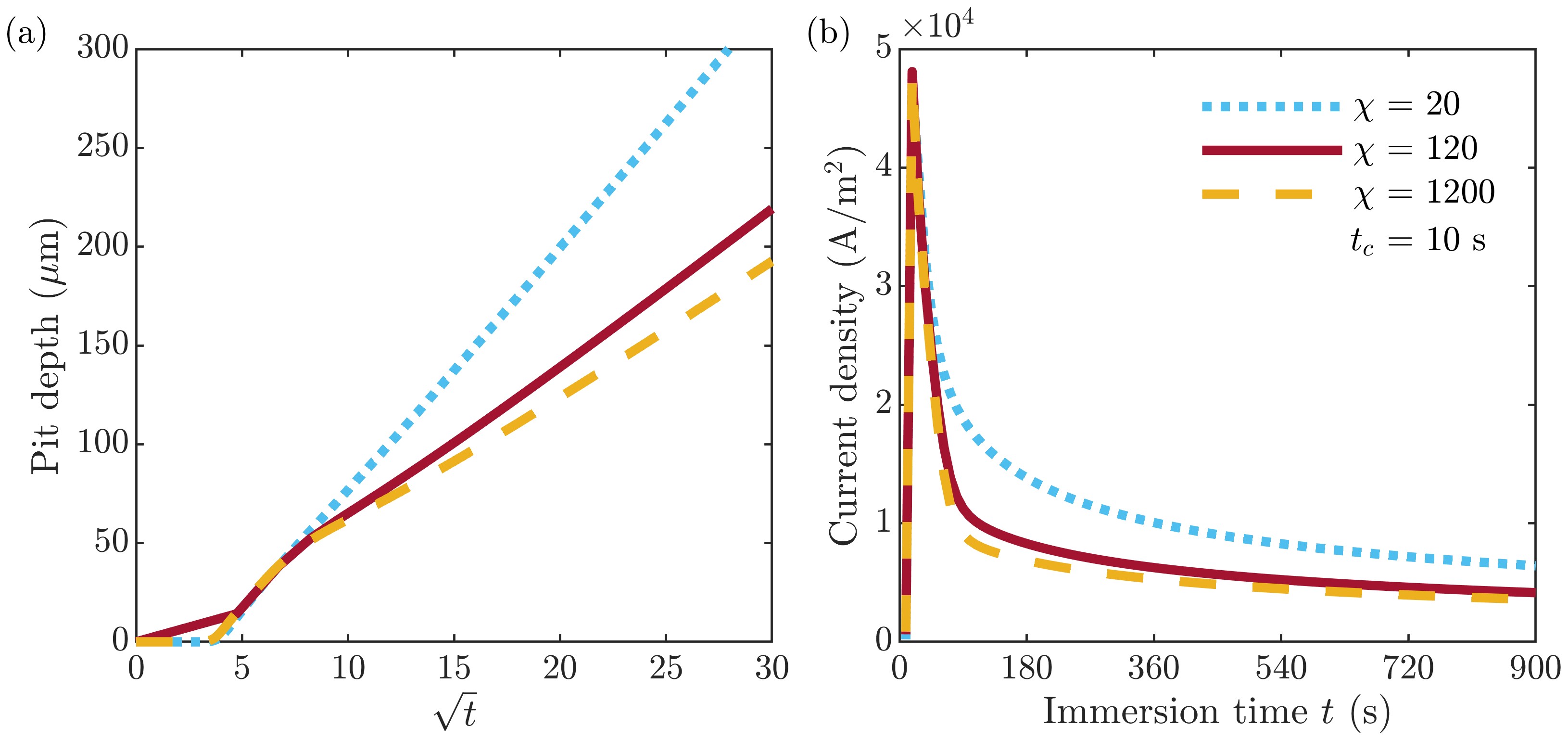}
    \captionsetup{labelfont = bf,justification = raggedright}
    \caption{Effect of the resistance proportionality constant $\chi$ on (a) the evolution of pit depth and (b) current density as a function of immersion time. The ECM parameter $t_c = 10$ s is kept constant.}
    \label{Fig5}
\end{figure}

\begin{figure}[h!]
    \centering
    \includegraphics[width = 15cm]{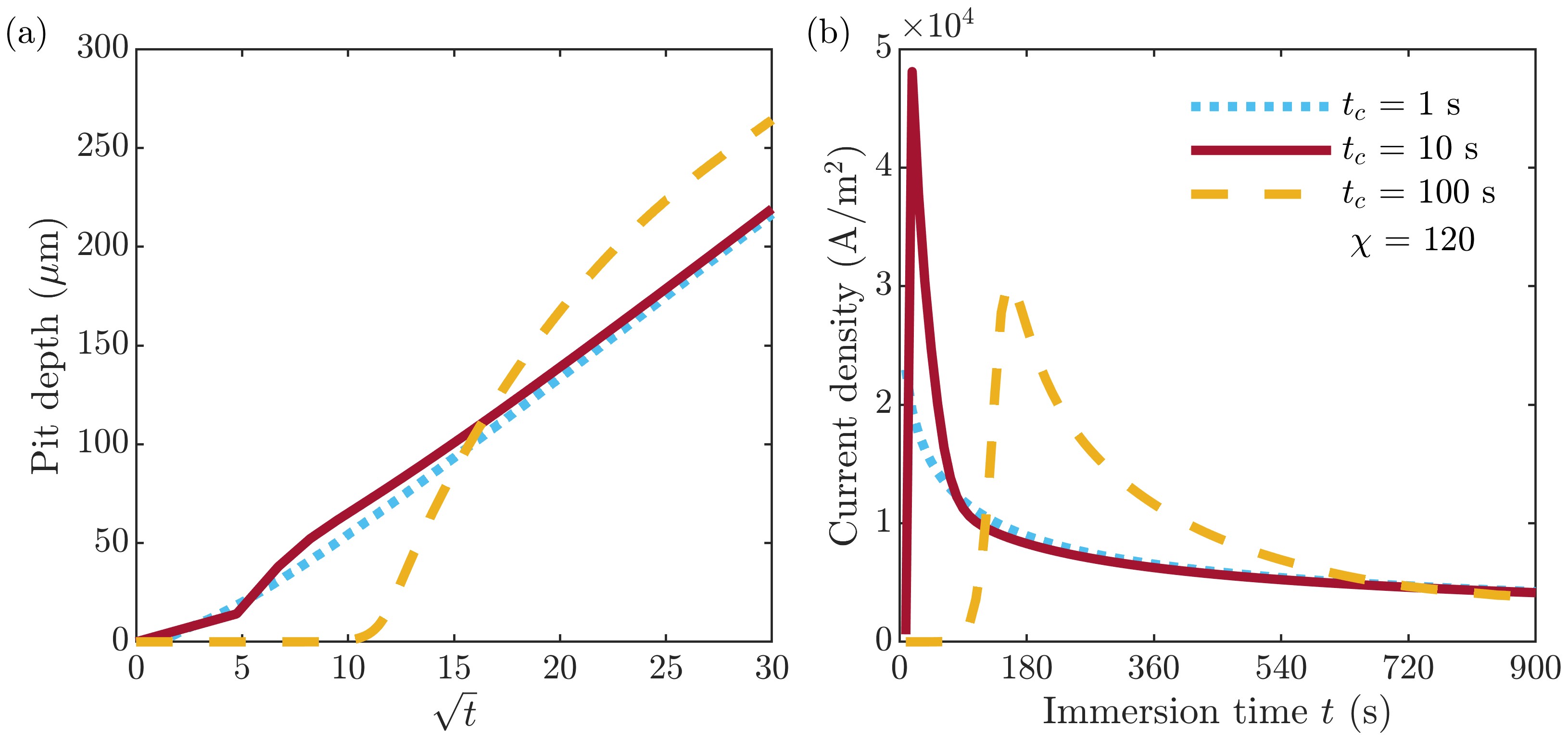}
    \captionsetup{labelfont = bf,justification = raggedright}
    \caption{Effect of the half-time of capacitor charging $t_c$ on (a) the evolution of pit depth and (b) current density as a function of immersion time. The ECM parameter $\chi = 120$ is kept constant.}
    \label{Fig6}
\end{figure}

The model predictions in terms of pit depth and current density for diffusion$-$controlled corrosion and different values of $\chi$ and $t_c$ are given in figure \ref{Fig5} and figure \ref{Fig6}. The diffusion$-$controlled process is achieved by utilizing the same applied potential as in the pencil electrode test in the previous section (600 mV (vs. SCE)). Figure \ref{Fig5} shows that decreasing $\chi$ increases pit depth and current density. This result follows from the fact that reducing $\chi$ leads to an increase in solution potential at the metal$-$electrolyte interface (equation (\ref{eqn18})), resulting in faster transport of metal ions far from the interface due to the increase in electromigration. This returns a faster dissolution rate and higher current density. The sensitivity of the parameter $t_c$ on the pit depth and current density is given in figure \ref{Fig6}. The obtained results show that the parameter $t_c$ determines the time at which the pit propagates and the time at which the peak of current density occurs. For the shortest $t_c$ considered ($t_c = 1$ s), the growth of the pit is triggered immediately and the peak of current density occurs at very short immersion times. For the longest $t_c$ considered ($t_c = 100$ s), the propagation of the pit and the peak of current density are postponed. The observed behavior is related to the solution potential at the metal$-$electrolyte interface and the strength of electromigration. The magnitude of solution potential at the interface is governed by equation (\ref{eqn18}). Higher solution potentials are returned for longer values of $t_c$ and cancel surface polarization in equation (\ref{eqn13}), decreasing overpotential. Such a reduced overpotential yields lower interface mobility (equation (\ref{eqn14})), which prolongs pit growth initiation.

\subsection{Distribution of ionic species and solution potential} \label{sec33}

\begin{figure}[h!]
    \centering
    \includegraphics[width = 15.0cm]{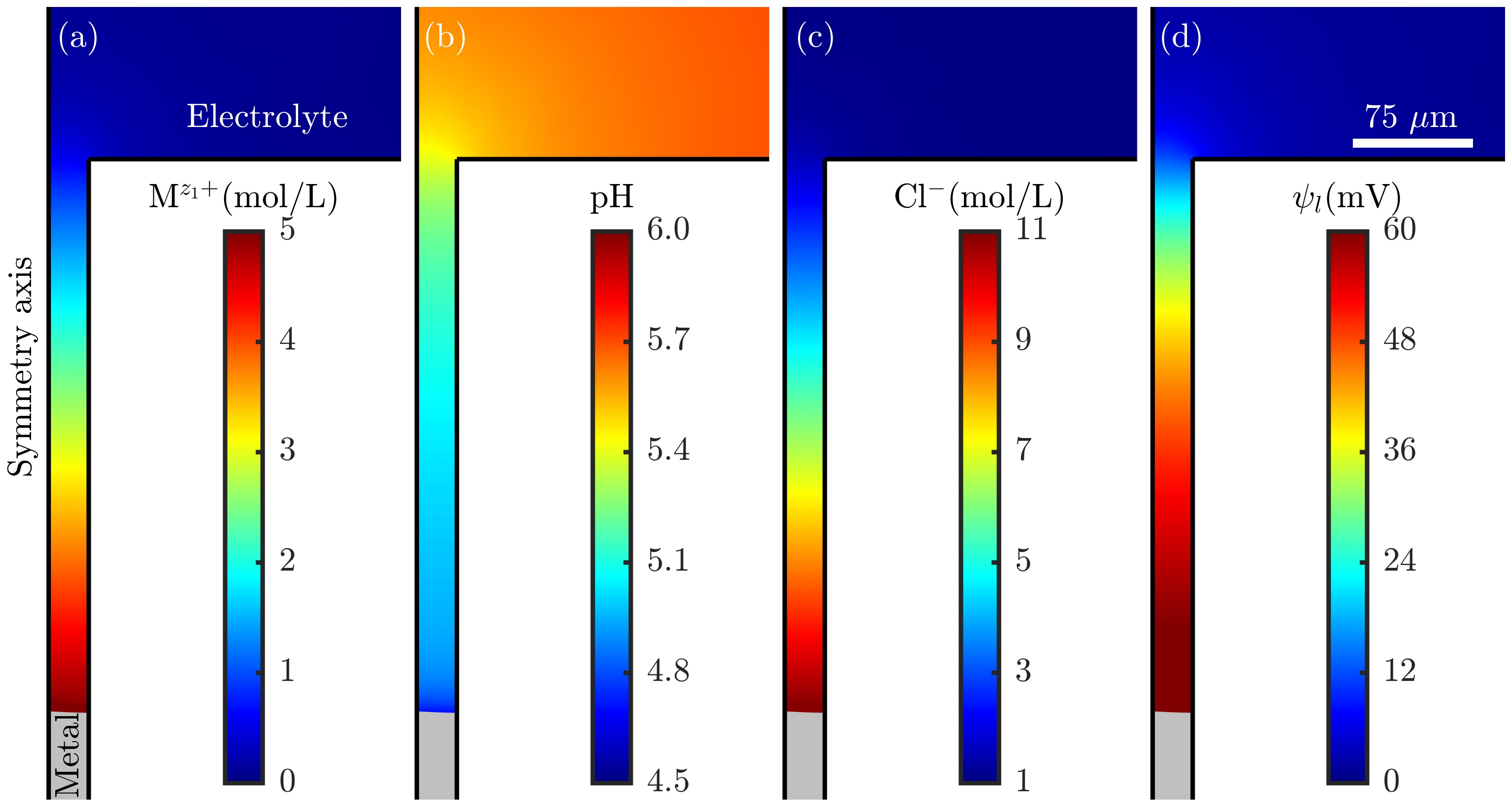}
    \captionsetup{labelfont = bf,justification = raggedright}
    \caption{Distribution of (a) metal ions $\text{M}^{z_1+}$, (b) pH, (c) chloride ions Cl$^-$, and (d) solution potential (vs. SCE). The resistance ratio is $\chi$ = 20 and the charging half-time is $t_c$ = 10 s. The grey area indicates the remaining undissolved electrode. The legend bars apply to the whole computational domain (figure \ref{Fig2}).}
    \label{Fig7}
\end{figure}

\begin{figure}[h!]
    \centering
    \includegraphics[width = 15.0cm]{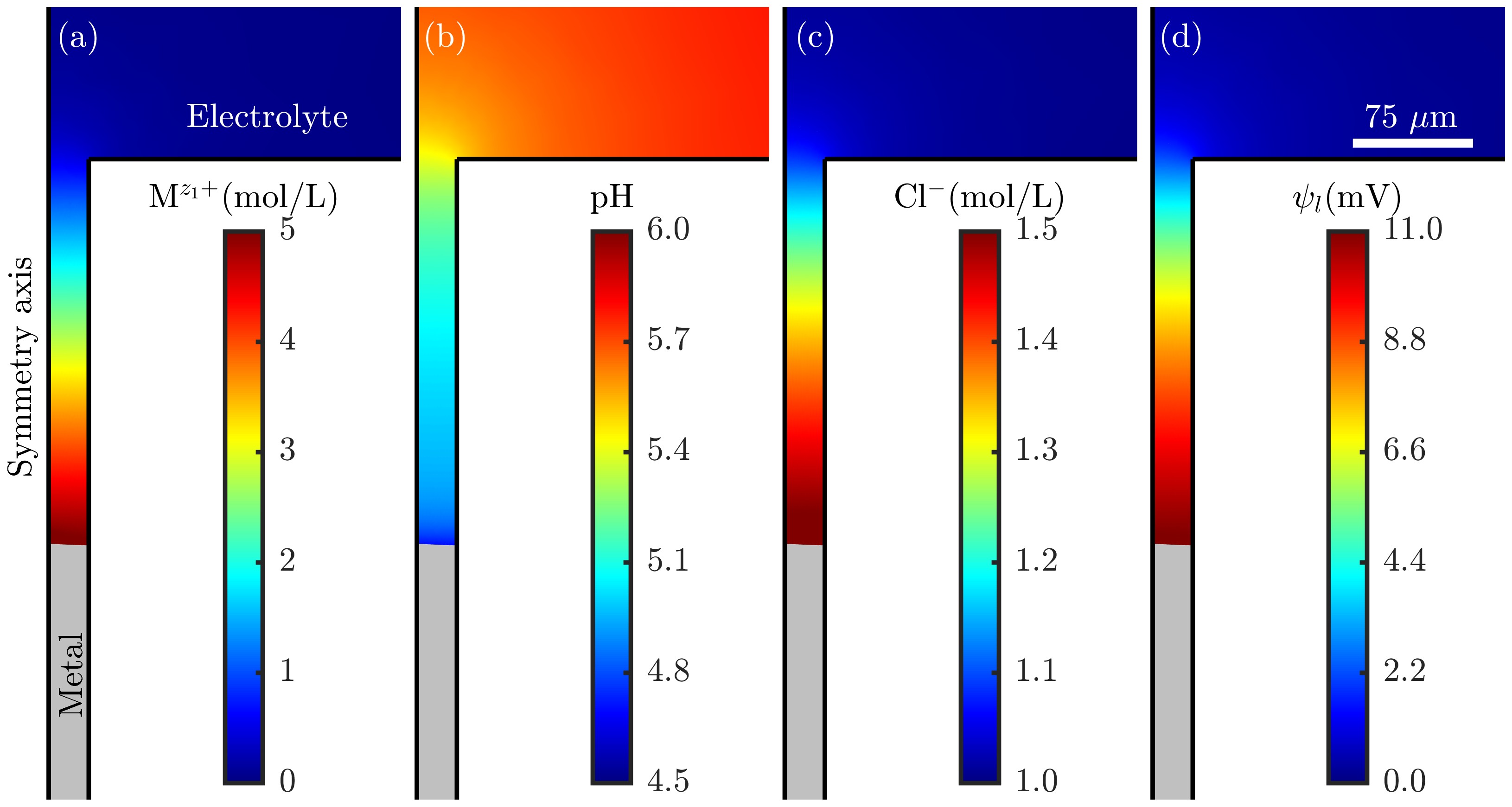}
    \captionsetup{labelfont = bf,justification = raggedright}
    \caption{Distribution of (a) metal ions $\text{M}^{z_1+}$, (b) pH, (c) chloride ions Cl$^-$, and (d) solution potential (vs. SCE). The resistance ratio is $\chi$ = 120 and the charging half-time is $t_c$ = 10 s. The grey area indicates the remaining undissolved electrode. The legend bars apply to the whole computational domain (figure \ref{Fig2}).}
    \label{Fig8}
\end{figure}

\begin{figure}[h!]
    \centering
    \includegraphics[width = 15.0cm]{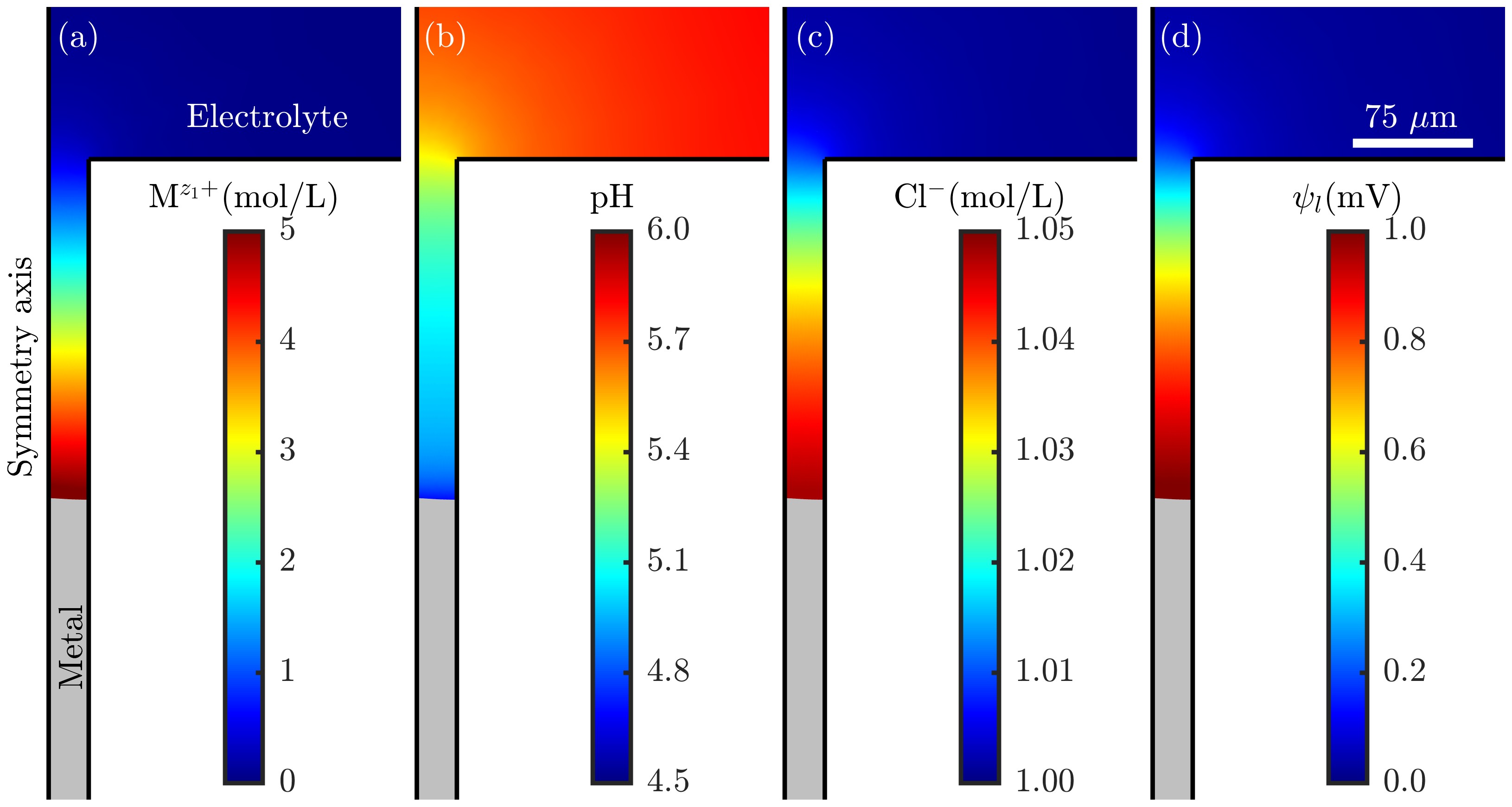}
    \captionsetup{labelfont = bf,justification = raggedright}
    \caption{Distribution of (a) metal ions $\text{M}^{z_1+}$, (b) pH, (c) chloride ions Cl$^-$, and (d) solution potential (vs. SCE). The resistance ratio is $\chi$ = 1200 and the charging half-time is $t_c$ = 10 s. The grey area indicates the remaining undissolved electrode. The legend bars apply to the whole computational domain (figure \ref{Fig2}).}
    \label{Fig9}
\end{figure}

The sensitivity of the ECM parameters $\chi$ and $t_c$ on the environmental response in terms of the distribution of ionic species and solution potential is presented in this section. The same material constants, numerical parameters, and applied potential (600 mV (vs. SCE)) used in Section \ref{sec31} are employed here. Figure \ref{Fig7}, figure \ref{Fig8}, and figure \ref{Fig9} display distributions of metal ions $\text{M}^{z_1+}$, pH, Cl$^-$, and electric potential at the final computational time. Figure \ref{Fig7} presents the case with the strongest electromigration using $\chi$ = 20 and $t_c = 10$ s. The obtained results for the weakest electromigration using $\chi$ = 1200 and $t_c$ = 10 s are shown in figure \ref{Fig9}. The distribution of ionic species and solution potential for the reference case used in Section \ref{sec31} for model calibration ($\chi$ = 120 and $t_c = 10$ s) is presented in figure \ref{Fig8}. The difference in dissolved electrode between figure \ref{Fig7}, figure \ref{Fig8}, and figure \ref{Fig9} indicates different dissolution rates. The concentration of metal ions at the metal$-$electrolyte interface reaches the equilibrium concentration in the liquid phase $c_{1}^{l,eq}$ for all three cases considered. Once the equilibrium concentration is reached, the difference in dissolution only arises from ionic migration far from the interface. The electric potential distribution in figure \ref{Fig7}(d), figure \ref{Fig8}(d), and figure \ref{Fig9}(d) indicates its contribution to ionic migration. The larger gradient of electric potential in figure \ref{Fig7}(d) produces a much more substantial contribution to electromigration of ions than those of figure \ref{Fig8}(d) and figure \ref{Fig9}(d). Notice that the maximum solution potential at the interface for these three cases considered is 60 mV, 11 mV, and 1 mV (vs. SCE). The high solution potential of 60 mV (vs. SCE) results in the fast removal of metal ions from near the interface (figure \ref{Fig7}(a)), leading to a fast dissolution rate.


The solution potential through electromigration also influences the transport of other ions present in the electrolyte. Chloride ions Cl$^-$ do not participate in the reactions listed in equations (\ref{eqn1}), (\ref{eqn2}), and (\ref{eqn3}). The competition between diffusion and electromigration (equation (\ref{eqn15})) governs the transport of Cl$^-$ ions in the electrolyte. Although the simulations start from the same uniform initial concentration of Cl$^-$ ions, different values of Cl$^-$ ions are returned at the metal$-$electrolyte interface. A comparison between the maximum value of solution potential and chloride ions in figure \ref{Fig7}, figure \ref{Fig8}, and figure \ref{Fig9} demonstrates the effect of solution potential in governing the transport of Cl$^-$ ions. The same observation can be noticed for the other ions. The local concentration of metal ions $\text{M}^{z_1 +}$, and their proportion to metal hydroxide $\text{M(OH)}^{(z_1 - 1)+}$, controls the maximum concentration of hydrogen ions $\text{H}^+$. For large enough backward reaction constants, $k_{1b}$ and $k_{2b}$, the maximum concentration of $\text{H}^+$ ions remains unchanged regardless of electromigration. Hydrogen, as positively charged, migrates to the bulk electrolyte along the solution potential gradient, increasing its acidity.

\begin{figure}[h!]
    \centering
    \includegraphics[width = 15cm]{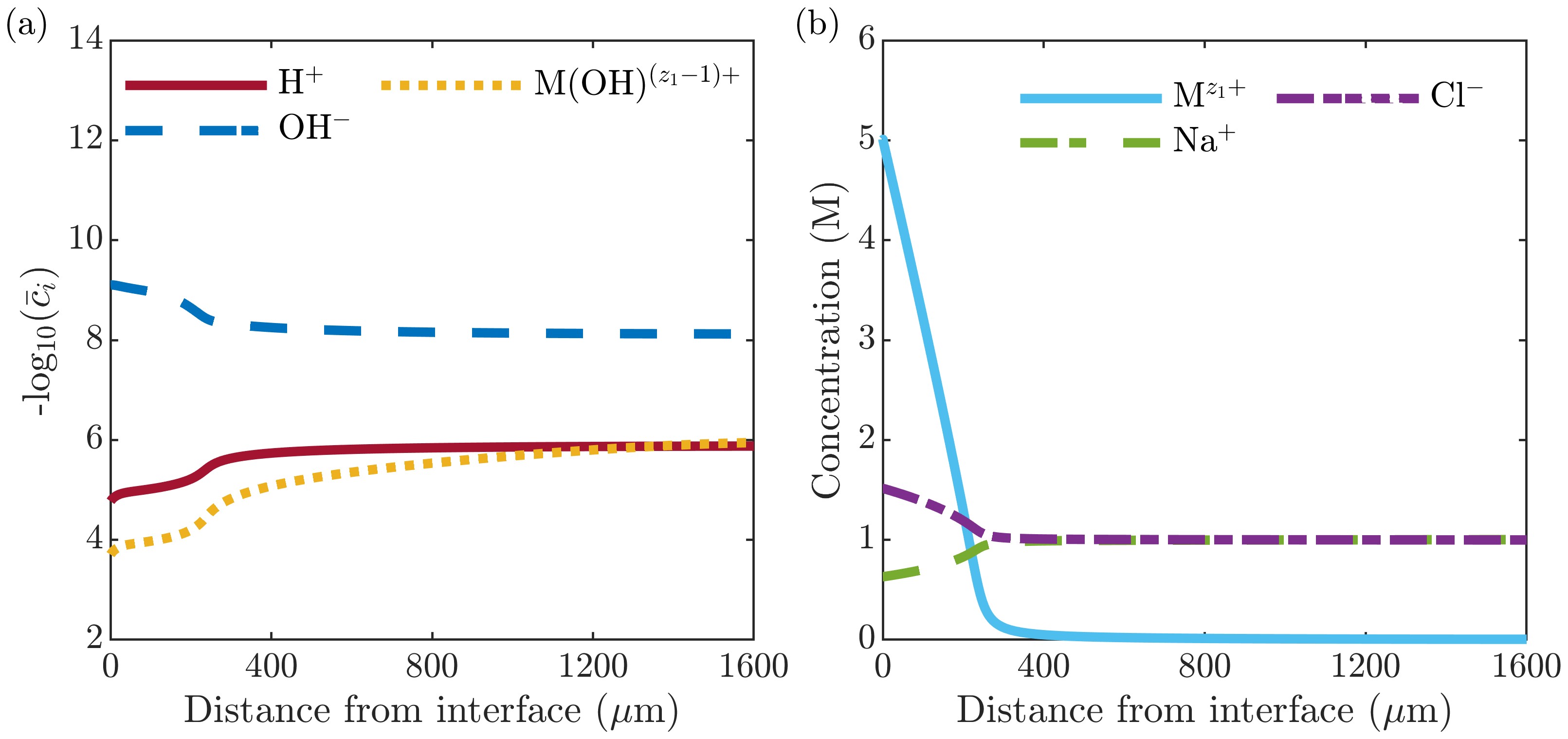}
    \captionsetup{labelfont = bf,justification = raggedright}
    \caption{Distribution of concentrations along the symmetry axis (figure \ref{Fig2}) for (a)  H$^+$, OH$^-$, M(OH)$^{(z_1 - 1)+}$ and (b) M$^{z_1 +}$, Na$^+$, and Cl$^-$ ions at the final computational time. $\bar{c}_i$ in (a) denotes the concentration of ions normalized with 1 mol/L. The ECM parameters are set to $\chi$ = 120 and $t_c$ = 10 s.}
    \label{Fig10}
\end{figure}

The distribution of ionic species along the symmetry axis (figure \ref{Fig2}) for the reference case study ($\chi=120$ and $t_c$ = 10 s) is shown in figure \ref{Fig10}. Both pH and metal ion range agree with figure \ref{Fig8}. Chloride ion distribution increases up to 1.5 M near the metal$-$electrolyte interface, decaying to an initial condition value of 1 M in the bulk electrolyte. Figure \ref{Fig10} also includes ionic species absent in figure \ref{Fig8}. As can be observed in figure \ref{Fig10}(a), hydroxide anions decrease with an increase in hydrogen cations. The mutual influence of dissociated water components comes from the water stability equation (\ref{eqn3}). It enforces that the sum of pH and pOH is constant and equal to 14. This verifies that the selected values of the penalty constants $k_{1b} = 2000$ m$^3$/(mol s) and $k_{2b} = 10000$ m$^3$/(mol s) (Appendix C) are large enough to satisfy the equilibrium conditions in equation (\ref{eqn16}) in the entire electrolyte. The $\text{M(OH)}^{(z_1 - 1)+}$ ions have a high concentration in the occluded zone, followed by a continuous decrease in the distance in the bulk electrolyte. The distribution of $\text{M}^{z_1 +}$, Na$^+$, and Cl$^-$ ions are shown in figure \ref{Fig10}(b). The distribution of Na$^+$ and Cl$^-$ ions is only affected by electromigration as they do not contribute to any reaction considered. There is a decrease in Na$^+$ ion concentration in the occluded zone, which then increases to an initial condition concentration of 1 M in the bulk electrolyte. The opposite behaviour is noticed for Cl$^-$ ions as they are negatively charged. As can be seen in figure \ref{Fig7}, figure \ref{Fig8}, figure \ref{Fig9}, and figure \ref{Fig10}, the localized processes occur inside the occluded zone while the composition of the bulk electrolyte does not evolve significantly from the initial conditions. All the ionic concentrations remain close to the initial conditions in the bulk electrolyte.

\begin{figure}[h!]
    \centering
    \includegraphics[width = 15.0cm]{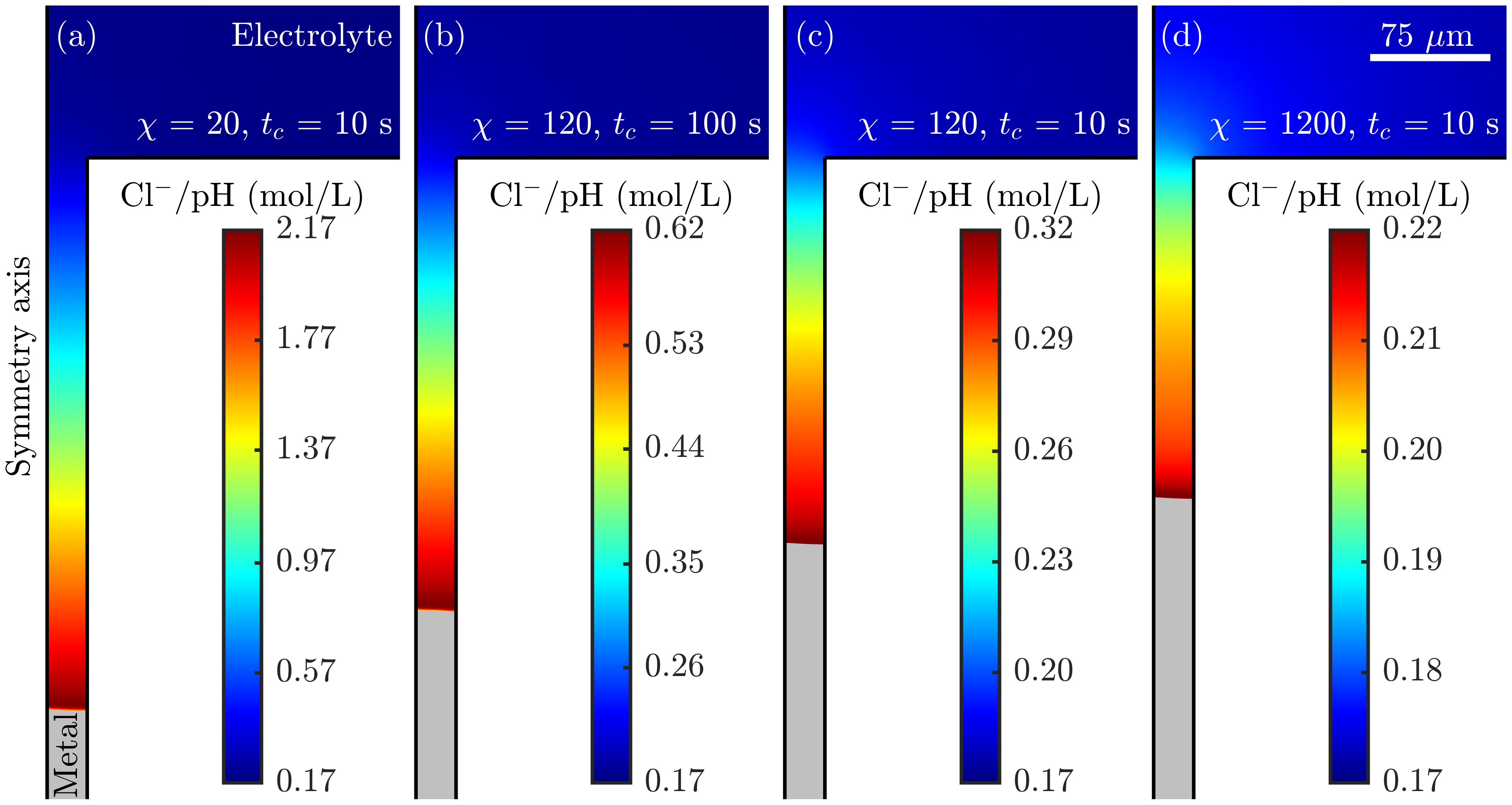}
    \captionsetup{labelfont = bf,justification = raggedright}
    \caption{The ratio between chloride ions and pH inside the occluded zone for (a) $\chi$ = 20, $t_c$ = 10 s, (b) $\chi$ = 120, $t_c$ = 100 s, (c) $\chi$ = 120, $t_c$ = 10 s, and (d) $\chi$ = 1200, $t_c$ = 10 s. The most severe attack corresponds to the highest ratio between Cl$^-$ and pH.}
    \label{Fig11}
\end{figure}

The obtained results for different ECM parameters $\chi$ and $t_c$ are used to determine the ratio between Cl$^-$ ions and pH inside the occluded zone. This ratio is commonly associated with damage in stainless steel \cite{MALIK1992, ERNST2002a, TIAN2015}, as it affects the stability of oxide films and precipitated phases \cite{Watanabe2022}. Four different cases are presented in figure \ref{Fig11}. The most severe attack corresponds to the highest ratio between Cl$^-$ and pH. It is obtained for the strongest electromigration ($\chi$ = 20 and $t_c = 10$ s), figure \ref{Fig11}(a). The second most damaging case is for $\chi$ = 120 and the longest charging kinetics of the EDL $t_c$ = 100 s, figure \ref{Fig11}(b). The ratio between Cl$^-$ and pH for the ECM parameters determined in Section \ref{sec31} for model calibration ($\chi$ = 120 and $t_c$ = 10 s) is shown in figure \ref{Fig11}(c). The case with the weakest electromigration, and thus, the least damaging case, is returned for $\chi$ = 1200 and $t_c = 10$ s, figure \ref{Fig11}(d). As can be observed in figure \ref{Fig11}, the ratio between chloride concentration and pH significantly depends on the ECM parameters $\chi$ and $t_c$. It mostly varies in the occluded zone and has a constant value in the bulk electrolyte for all the cases considered. Moreover, figure \ref{Fig11} shows a correlation between increased dissolution and increased concentration of harmful ionic species near the metal$-$electrolyte interface.

\section{Discussion} \label{sec4}

Tracking the metal$-$electrolyte interface and measuring pit kinetics in situ is challenging due to the small length scale and the formation of corrosion products. As such, the current density is typically reported in experimental studies to quantify corrosion damage. On the other hand, numerical models are mostly validated against pit kinetics. The present model reproduces experimental data on both pit depth and current density, figure \ref{Fig3}. The model developed enables more accurate information on corrosion damage and captures the current density response, providing a more robust framework for quantifying corrosion damage. This feature is enabled by solving a Laplace-type equation (\ref{eqn17}) for the electric potential along with the boundary condition (equation (\ref{eqn18})) prescribed on the metal surface $\partial\Omega_m$, figure \ref{Fig1}. The boundary condition is a function of ECM parameters $\chi$ and $t_c$, which define the properties of the EDL, such as its resistance and capacitance. The resulting solution potential distribution depends on the parameters $\chi$ and $t_c$, which can be estimated from experimental studies, as illustrated in figure \ref{Fig4}.

Alternative approaches for solving for solution potential are based on a Poisson-type equation
\begin{equation} \label{eqn19}
    - \nabla \cdot \big(\lambda \nabla \psi_l \big) = \zeta \quad \text{in}\quad\Omega; \quad\mathrm{} \lambda \mathbf{n}\cdot \nabla \psi_l = 0 \quad \text{on}\quad\partial\Omega,
\end{equation}
where $\zeta$ is the source term introduced to represent the net change in charge density as a result of the electrochemical reaction. The source term $\zeta$ is commonly expressed considering the production of electrons due to the dissolution of the metal electrode \cite{Tsuyuki2018, LIN2019, Tantratian2022, CUI2023}
\begin{equation} \label{eqn20}
    \zeta =  F c_{1}^{s,eq} z_1 \frac{\partial \phi}{\partial t},
\end{equation} 
or assuming that the conservation of charge is at a steady state \cite{Ansari2018, LIN2020, Chen2022}
\begin{equation} \label{eqn21}
    \zeta =  F \sum z_i \frac{\partial c_i}{\partial t}.
\end{equation}
Similar alternative expressions for the source term $\zeta$ can be found in \cite{Chadwick2018, Mai2018}. The solution potential distribution determined using equation (\ref{eqn19}) depends on the rate of interface evolution or the rate of ionic species evolution. The obtained solution potential is fed back in the electrochemical flux $\mathbf{J}_i$ in the mass transport equation (\ref{eqn15}). This two-way coupling between the solution potential and phase-field variable or ionic species can lead to excessive electromigration. Such an increase in electromigration enhances the transport of ions away from the interface, further increasing dissolution. Moreover, it is challenging with this approach to attain the expected linear trend in pit kinetics and immersion time (figure \ref{Fig3}(a)), as demonstrated below.

\begin{figure}[h!]
    \centering
    \includegraphics[width = 8cm]{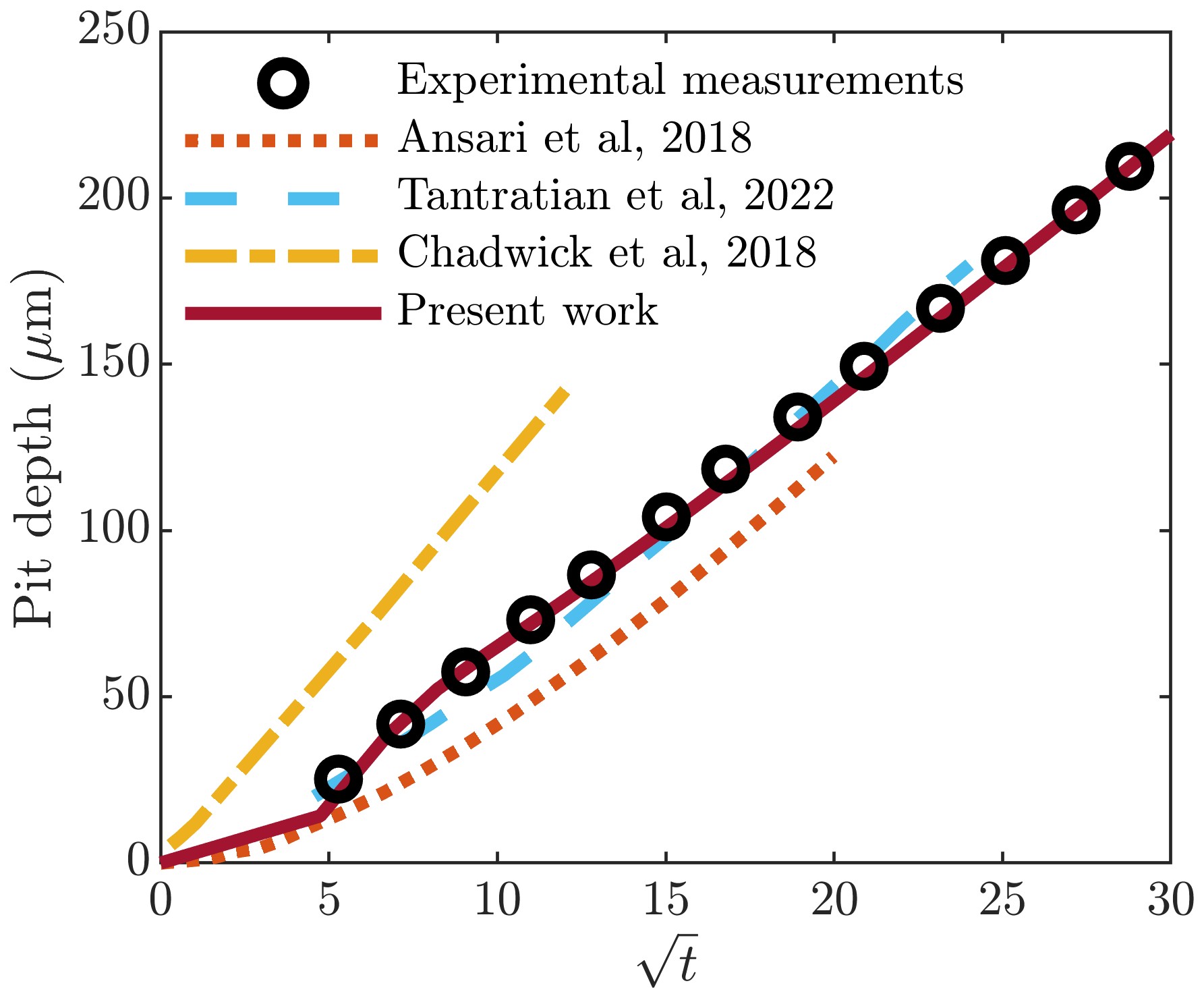}
    \captionsetup{labelfont = bf,justification = raggedright}
    \caption{Comparison between the present work and models from the literature (Ansari et al. \cite{Ansari2018}, Tantratian et al. \cite{Tantratian2022}, Chadwick et al. \cite{Chadwick2018}) for the evolution of pit depth as a function of immersion time. Experimental measurements are taken from Ref. \cite{ERNST2002a}.}
    \label{Fig12}
\end{figure}

A comparison between the present work and representative studies that utilize equation (\ref{eqn19}), along with a different source term $\zeta$, to solve for solution potential is shown in figure \ref{Fig12}. The same pencil electrode test and experimental measurements of pit depth used in Section \ref{sec31} for calibration of the present model are used in Refs. \cite{Ansari2018, Tantratian2022, Chadwick2018}. The comparison is performed against pit depth as the original references \cite{Ansari2018, Tantratian2022, Chadwick2018} do not report transient current density response. As can be observed in figure \ref{Fig12}, these studies underestimate pit depth \cite{Ansari2018}, cover a limited range of experimental data \cite{Tantratian2022} or overestimate the pit depth kinetics \cite{Chadwick2018}. Moreover, they are unable to reproduce the linear trend in pit kinetics and the square root of immersion time. The present work satisfactorily covers the whole range of experimental data and is in agreement with experimental measurements. The solution potential distribution in the present model does not depend on the rate of interface evolution or the rate of ionic species evolution. The boundary condition in equation (\ref{eqn18}), which is a function of ECM parameters $\chi$ and $t_c$, controls the maximum solution potential drop across the electrolyte. The ionic species are no longer subjected to excessive electromigration. The resulting electromigration flux enhances the ionic migration enough for the pit depth propagation to meet the experimental data. The comparison in figure \ref{Fig12} highlights the necessity of introducing charging dynamics of the EDL to describe damage propagation accurately in accelerated corrosion tests. 

\begin{figure}[h!]
    \centering
    \includegraphics[width = 15 cm]{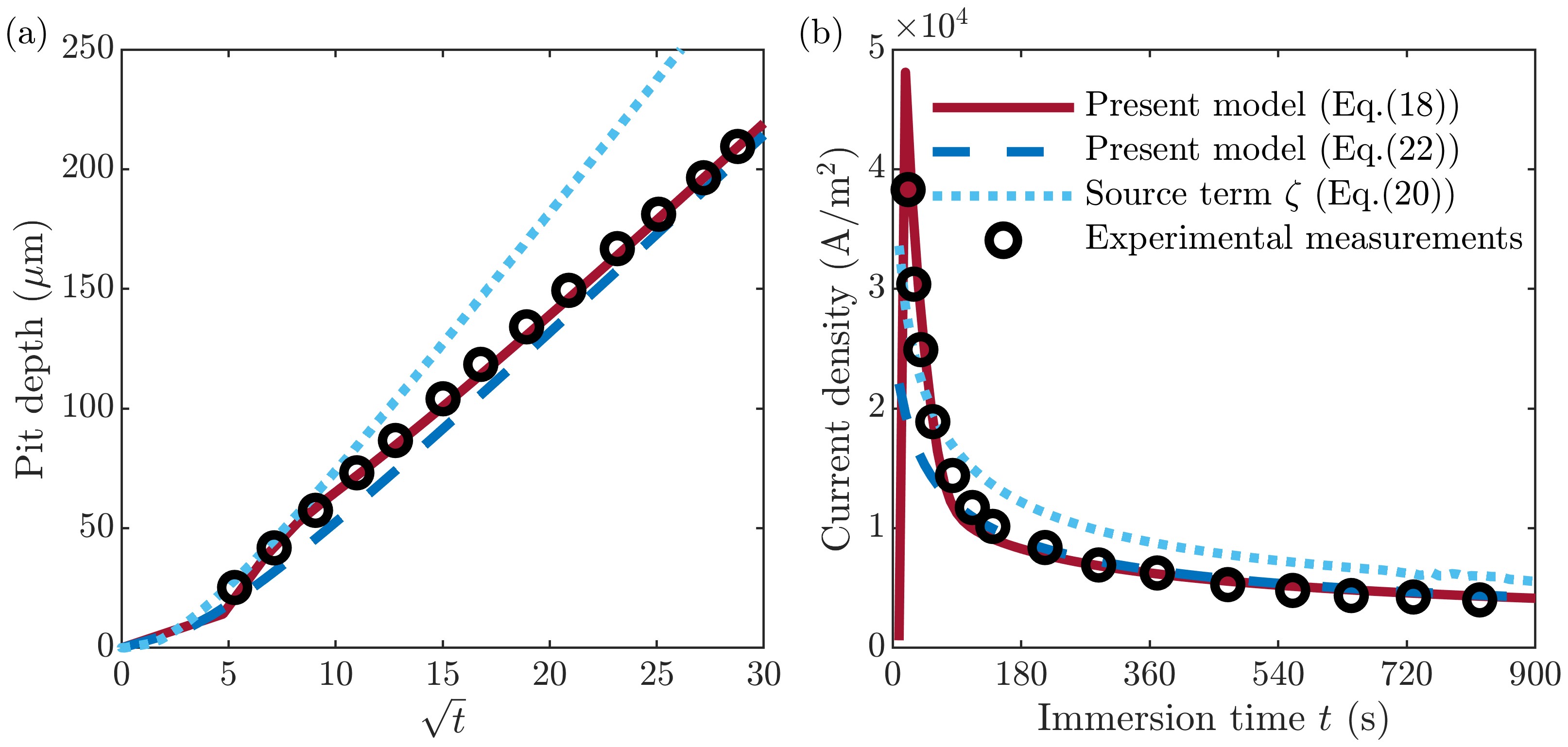}
    \captionsetup{labelfont = bf,justification = raggedright}
    \caption{Comparison between experimental measurements \cite{ERNST2002a} and predictions of different models of (a) the evolution of pit depth and (b) current density as a function of immersion time.}
    \label{Fig13}
\end{figure}

Additional analysis is performed to show the effect of solution potential distribution on the pit depth and current density response. Three cases are considered in this study. The first case includes the present model in which the solution potential is obtained using equation (\ref{eqn17}) and boundary condition (\ref{eqn18}). The second case considers the same model but neglects the exponential term in equation (\ref{eqn18}). This case follows from the sensitivity study in Section \ref{sec32}, which reveals that this term can be ignored for a long immersion time $t \gg t_c$. The solution potential distribution still follows the same governing equation (\ref{eqn17}) and is subjected to a constant boundary condition
\begin{equation}\label{eqn22}
\psi_l^{dl} = \psi^0 \frac{1}{1 + \chi} \quad \text{on}\quad\partial\Omega_m.
\end{equation}
The solution potential distribution in the third case considered is obtained utilizing equation (\ref{eqn19}) and the source term $\zeta$ given in equation (\ref{eqn20}). This case study introduces a dependence of solution potential on the rate of interface evolution. The same material properties and numerical constants given in Table \ref{table1} are employed in this study. A comparison between these three models considered in terms of pit depth and current density is given in figure \ref{Fig13}. As shown in figure \ref{Fig13}, the model that uses the source term in the governing equation for the solution potential overestimates the experimental data. The discrepancy arises from excessive electromigration contribution. Such an increase in electromigration flux is responsible for the faster motion of ions far from the interface, returning an accelerated dissolution rate and higher current density. The pit kinetics and current density response are similar between the present models that utilise equation (\ref{eqn18}) and equation (\ref{eqn22}) for the boundary condition in the governing equation for the solution potential. Models that neglect charging dynamics of the EDL (equation (\ref{eqn22})) slightly underpredict pit depth and current density response at early immersion times. Afterwards, pit depth and current density align with the experimental data, illustrating that the capacitive process has a limited influence at longer immersion times.

The main disadvantage of the present model is the number of additional parameters that need to be calibrated. When dissolution is a diffusion-controlled process, the kinetic coefficient $L_0$ is a purely numerical parameter that needs to be calibrated against experimental data. However, it has been shown that for activation-controlled dissolution, the kinetic coefficient $L_0$ is directly proportional to anodic current density \cite{CUI2022}. The ECM parameters $\chi$ and $t_c$ can be calibrated against current density data or extracted from experimental studies, as evidenced in figure \ref{Fig4}. A single pair of $\chi$ and $t_c$ describes a specific system. On the other hand, incorporating the ECM parameters into the present framework decouples the solution potential evolution from the rate of interface evolution or the rate of ionic species evolution. This decoupling eliminates excessive electromigration, enabling more accurate information on corrosion damage by capturing both pit kinetics and current density. Matching current density response throughout the charging of the EDL is a promising step towards correlating natural and accelerated corrosion tests. The main approximations used in the present investigation are related to the chemical free energy density in Section {\ref{sec22}}. The chemical free energy density associated with the metal ion concentration in both phases is approximated with parabolic functions with the same density curvature parameter $A$, equation ({\ref{eqn7}}). As such, it assumes that the metal and the electrolyte are solutions with considerable composition ranges. Moreover, the dilute solution theory is used to express the contribution to the chemical free energy density from the other ions present in the electrolyte, equation ({\ref{eqn9}}). A more detailed description of the chemical free energy density of the system is required to verify those assumptions. This should be addressed in future work. It is also assumed in the present study that the ECM parameters $\chi$ and $t_c$ are constants and independent of the concentration of ionic species near the metal$-$electrolyte interface. While this approximation is justified for short accelerated corrosion tests, as the one considered in Section {\ref{sec31}}, it may not be valid for longer immersion times, as shown in figure {\ref{Fig4}}. As demonstrated in Section {\ref{sec32}} and Section {\ref{sec33}}, the variation in the ECM parameters influences pit depth, current density, and the distribution of ionic species and solution potential in the electrolyte. Hence, they should be chosen judiciously.

\section{Conclusions} \label{sec5}

A computational framework based on the phase-field method is developed to assess the corrosion of metallic materials. The model incorporates the role of electrochemistry and charging kinetics of the EDL in governing the corrosion process. The effects of the EDL on the transport of ionic species in the electrolyte are established by introducing a boundary condition on the solution potential equation. The properties of the EDL, such as its resistance and capacitance, are included in the model through two ECM parameters. The study shows that the ECM parameters determine the magnitude of solution potential at the electrode$-$electrolyte interface, which in turn, defines the magnitude of electromigration. As a result, different pit kinetics, current density, and environmental responses in terms of the distribution of ionic species and solution potential are obtained.

Future work should extend beyond potentiostatic tests and consider the dependence of the Pourbaix diagram on surface polarization. Such a model would allow capturing current density in electrochemical impedance spectroscopy and polarization studies. Future work should also consider the role of precipitation of stable phases such as salts on dissolution kinetics. The formation of an oxide layer can also be incorporated into the present framework following equivalent circuit models.

\section{Acknowledgments} \label{sec6}

M.M., M.R.W. and E.M.-P. acknowledge support from the EPSRC Centre for Doctoral Training in Nuclear Energy Futures [Grant EP/5023844/1]. S.K. and E.M.-P. acknowledge financial support from UKRI’s Future Leaders Fellowship program [Grant MR/V024124/1]. 

\section*{Data Availability} \label{sec7}

The code developed is available at \url{https://mechmat.web.ox.ac.uk/codes}.

\section*{Appendix A. Dependence of electrode kinetics on overpotential} \label{appendixA}

The enhanced definition of interfacial mobility $L$ in equation (\ref{eqn14}) captures a nonlinear dependence of electrode kinetics on overpotential. This dependence on overpotential is demonstrated in figure \ref{FigA1} considering interface velocity and current density. Both quantities are determined using the present model and Butler$-$Volmer kinetics. As can be observed in figure \ref{FigA1}, the present model returns the trend predicted by Butler$-$Volmer equation (\ref{eqn13}). The present framework with this definition of interfacial mobility $L$ resembles nonlinear phase-field models of corrosion \cite{Chadwick2018, LIN2021, Brewick2022, Tantratian2022} formulated on reaction rate theory \cite{Linyun2012, Liang2014, CHEN2015a}.

\renewcommand{\thefigure}{A.1}
\begin{figure}[H]
    \centering
    \includegraphics[width = 15cm]{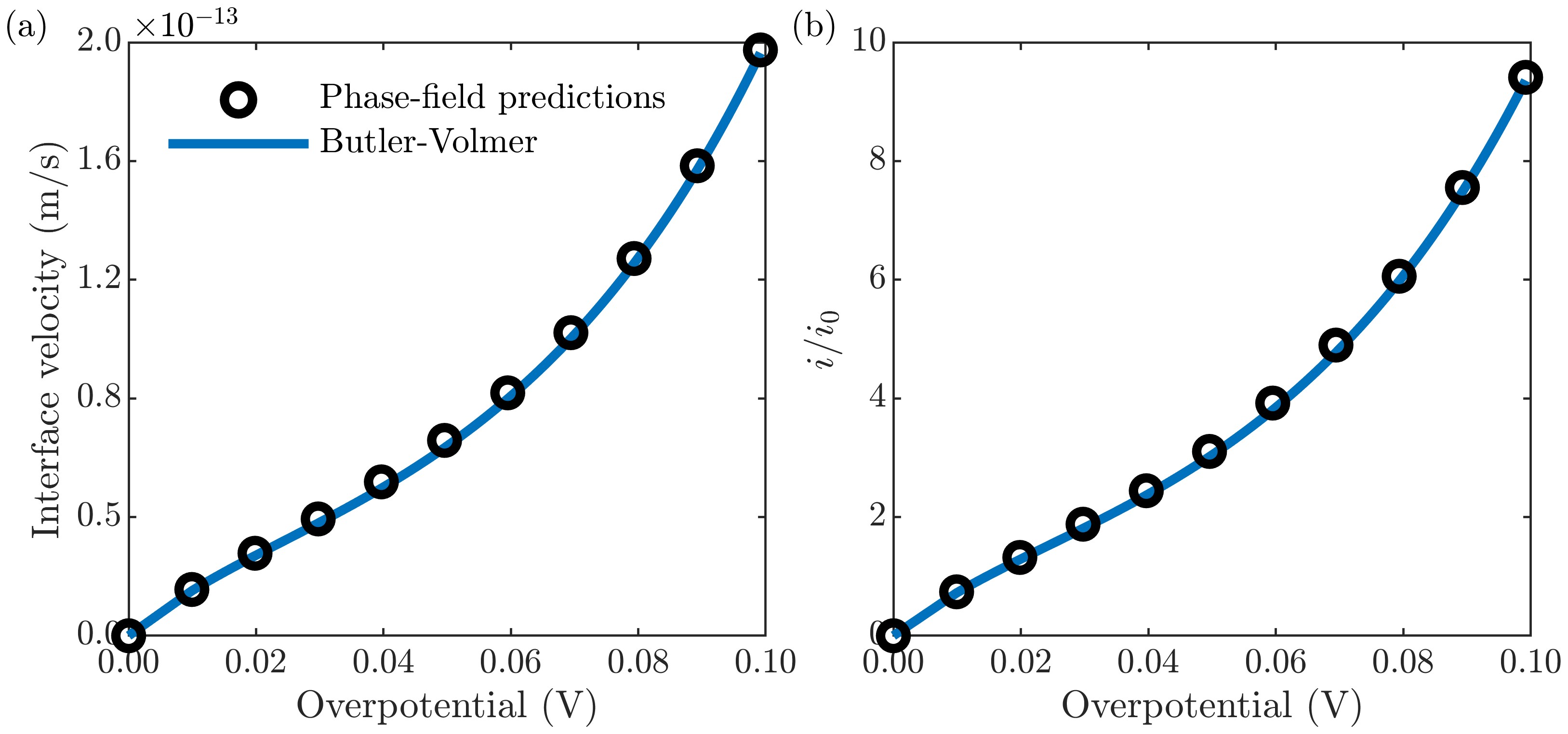}
    \captionsetup{labelfont = bf,justification = raggedright}
    \caption{Dependence of (a) interface velocity and (b) current density on overpotential.}
    \label{FigA1}
\end{figure}

\section*{Appendix B. Equivalent circuit models} \label{appendixB}

Two ECMs shown in figure \ref{FigB1} are frequently used in experimental practice to determine the properties of electric double layer (EDL). The ECM in figure \ref{FigB1}(a) is employed in this investigation. The circuit model in figure \ref{FigB1}(b) allows for capturing more capacitive processes at the metal$-$electrolyte interface. Simulation results in Section \ref{sec31} demonstrate that the simplest circuit model (figure \ref{FigB1}(a)) is sufficient to capture experimental data on pit kinetics and current density, figure \ref{Fig3}. The evolution of the solution potential at the EDL$-$electrolyte interface $\psi_l^{dl}$ is given in equation (\ref{eqn18}) considering the circuit model in figure \ref{FigB1}(a). Details regarding its derivation and an equivalent equation for the circuit model in figure \ref{FigB1}(b) are given below.

\renewcommand{\thefigure}{B.1}
\begin{figure}[H]
    \centering
    \includegraphics[width = 15cm]{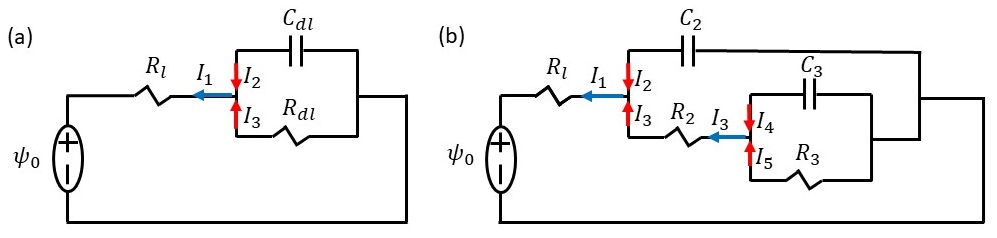}
    \captionsetup{labelfont = bf,justification = raggedright}
    \caption{Two equivalent circuit models most commonly used in experimental practice. Circuit model with (a) one and (b) two capacitive processes.}
    \label{FigB1}
\end{figure}

Following the first Kirchhoff Law, the conservation of currents at the contact between the bulk electrolyte and the EDL (figure \ref{FigB1}(a)) can be written as \cite{Sundararajan2020}
\begin{equation} \label{eqnA1} \tag{A.1}
     I_1 = I_2 + I_3 \quad\mathrm{}\quad I_1 = \frac{\psi^0 - \psi_l^{dl}}{R_l} \quad \quad  I_2 = C_{dl}\frac{d\psi_l^{dl}}{dt} \quad \quad I_3 = \frac{\psi_l^{dl}}{R_{dl}}, 
\end{equation}
where $\psi^0$ accounts for the initial surface polarization at time zero ($t=0$), $\psi_l^{dl}$ the evolution of the potential at the interface between the EDL and the electrolyte, $R_l$ the solution resistance, $R_{dl}$ the resistance of the EDL, and $C_{dl}$ the capacitance of the EDL. The resistor $R_{dl}$ and the capacitor $C_{dl}$ are connected in parallel and they experience the same electric potential drop. The conservation of currents is re-written as 
\begin{equation} \label{eqnA2} \tag{A.2}
    \frac{\psi^0 - \psi_l^{dl}}{R_l} = C_{dl}\frac{d\psi_l^{dl}}{dt} + \frac{\psi_l^{dl}}{R_{dl}}.
\end{equation}
Solving the previous expression for $\psi_l^{dl}$ leads to   
\begin{equation} \label{eqnA3}  \tag{A.3}
    \psi_l^{dl} = C \exp{ \bigg( -\frac{R_l + R_{dl}}{C_{dl}R_{l}R_{dl}}t \bigg)},
\end{equation}
where $C$ is the constant that determines the initial conditions for the discharge of the capacitor. Standard manipulation renders the following expression
\begin{equation} \label{eqnA4}  \tag{A.4}
   \psi_l^{dl} = \psi^0 \bigg( \frac{R_{l}}{R_{l} + R_{dl}} + \frac{R_{dl}}{R_{l} + R_{dl}}\exp{ \bigg( -\frac{R_l + R_{dl}}{C_{dl}R_{l}R_{dl}}t \bigg) } \bigg) .
\end{equation}
Upon assuming that the resistance of the EDL is proportional to the solution resistance ($R_{dl} = \chi R_l$) and that the capacitance of the EDL is formed at half-time of capacitor charging $t_{c}$ ($C_{dl} = \xi t_{c}/ (R_{dl} \text{ln}2)$), the evolution of the potential at the interface is written as
\begin{equation} \label{eqnA5}  \tag{A.5}
   \psi_l^{dl} = \psi^0 \bigg( \frac{1}{1 + \chi} + \frac{\chi}{1 + \chi}\exp{ \bigg( -\frac{t\ln{2}}{\xi t_c} \bigg) } \bigg),
\end{equation}
where $\xi$ stands for the geometric factor to account for the change in the electrode$–$electrolyte interfacial area during the dissolution process. The previous expression resembles equation (\ref{eqn18}).

Following the same steps as above, the evolution of the solution potential using the ECM in figure \ref{FigB1}(b) can be written as
\begin{equation} \label{eqnA6}  \tag{A.6}
    \psi_l^{dl} = \psi^0 \bigg( \frac{R_l}{R_l + R_2 + R_3} + \frac{R_2 + R_3}{R_l + R_2 + R_3} \exp{ \bigg( - \frac{R_{2}(R_{2}+R_{3})t+\frac{C_{3}R_{l}R_{2}R_{3}^{2}}{R_{2}+R_{3}}\exp{ \big( -\frac{R_2 + R_3}{C_{3}R_{2}R_3}t \big) }}{C_{2}R_{l}R_{2}(R_{2}+R_{3})} \bigg) } \bigg),
\end{equation}
where $R_2$ and $C_2$ are the resistance and capacitance of the outer capacitor, $R_3$ and $C_3$ are the resistance and capacitance of the inner capacitor. Assuming that the resistances are proportional ($\chi = R_2/R_l$, $\varphi = R_3/R_2$) and that the capacitance of each capacitor is formed at half-time of capacitor charging ($C_{2} = t_{c_2}/ (R_{2} \text{ln}2)$, $C_{3} = t_{c_3}/ (R_{3} \text{ln}2)$) renders the following expression for the evolution of the potential at the interface
\begin{equation} \label{eqnA7}  \tag{A.7}
     \psi_l^{dl} = \psi^0 \bigg( \frac{1}{1 + \chi (1+\varphi)} + \frac{\chi (1+\varphi)}{1+\chi (1+\varphi)} \exp{ \bigg( - \frac{ (1+\varphi)t\ln{2} + \frac{\varphi t_{c_3} }{\chi} \exp{ \big( -\frac{t\ln2}{t_{c_3}} \big) }}{(1+\varphi)t_{c_2} + \frac{\varphi t_{c_3}}{\chi \ln2}\exp{ \big( -\frac{t_{c_2}\ln2}{t_{c_3}} \big) }} \bigg) } \bigg).
\end{equation}
\cleardoublepage
\section*{Appendix C. Parametric sensitivity analysis for penalty constants $k_{1b}$ and $k_{2b}$} \label{appendixC}

The parametric sensitivity study is performed for the penalty constants $k_{1b}$ and $k_{2b}$. A convergence analysis is conducted in which the values of both constants are varied until the equilibrium conditions in equation ({\ref{eqn16}}) are not satisfied. As can be observed in figure {\ref{FigC1}}, the equilibrium conditions in equation ({\ref{eqn16}}) are met for $k_{1b}=2000$ m$^3$/(mol s) and $k_{2b} = 10000$ m$^3$/(mol s). A further increase in the values of the penalty constants leads to higher computational costs and convergence issues.   
 
\renewcommand{\thefigure}{C.1}
\begin{figure}[H]
    \centering
    \includegraphics[width = 15cm]{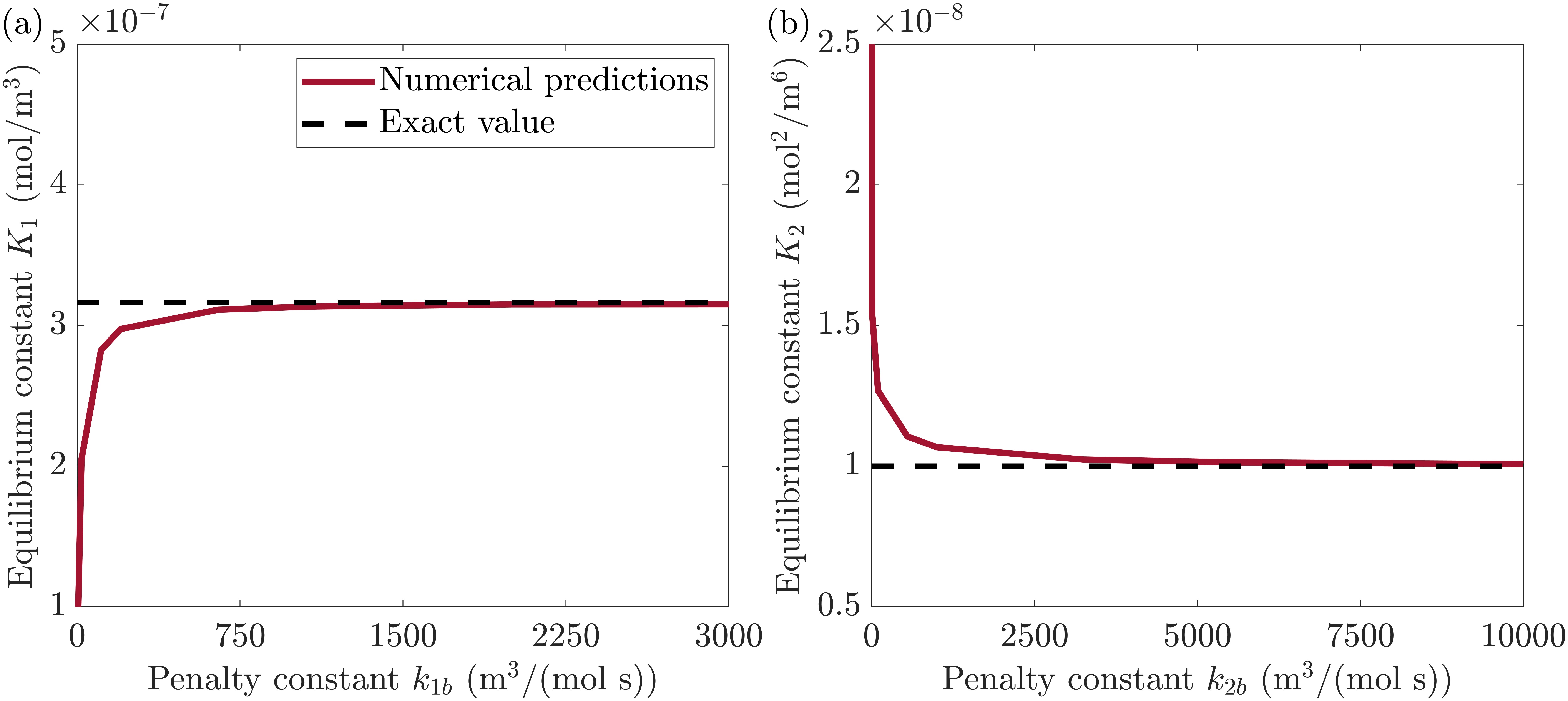}
    \captionsetup{labelfont = bf,justification = raggedright}
    \caption{Convergence analysis for the (a) equilibrium constant $K_1$ and (b) equilibrium constant $K_2$ as a function of penalty constants $k_{1b}$ and $k_{2b}$.}
    \label{FigC1}
\end{figure}

\begin{singlespace}

\end{singlespace}
\end{document}